\documentclass[aps,prd,12pt]{revtex4}
\begin{document}
\def\eth{\epsilon}
\def\emx{\varepsilon}
\def\ER{E_{\rm R}}
\def\SD{{\rm SD}}
\def\SI{{\rm SI}}
\renewcommand{\baselinestretch}{1.0}
\title{On dark matter search after DAMA with \boldmath $^{73}$Ge}     
\author{V.A.~Bednyakov} 
\affiliation{Laboratory of Nuclear Problems,
         Joint Institute for Nuclear Research, \\ 
         141980 Dubna, Russia; E-mail: Vadim.Bednyakov@jinr.ru} 
\author{and H.V.~Klapdor-Kleingrothaus} 
\affiliation{Max-Planck-Institut f\"{u}r Kernphysik, 
        Postfach 103980, D-69029, Heidelberg, Germany; 
	E-mail: H.Klapdor@mpi-hd.mpg.de} 
\begin{abstract} 
        The Weakly Interacting Massive Particle (WIMP)
        is one of the main candidates for the relic dark matter (DM).
	In the effective low-energy minimal supersymmetric standard
	model (effMSSM) 
	the neutralino-nucleon 
	spin and scalar cross sections in the low-mass regime were calculated. 
	The calculated cross sections are compared with almost all 
	experimental currently available exclusion curves 
        for spin-dependent WIMP-proton and WIMP-neutron cross sections. 
	It is demonstrated that {\it in general} 
        about two-orders-of-magnitude improvement 
        of the current DM experiment sensitivities is needed 
        to reach the (effMSSM) SUSY predictions. 
	At the current level
	of accuracy it looks reasonable to safely neglect 
	{\it sub-dominant spin WIMP-nucleon contributions}  
	analyzing the data from spin-non-zero targets.
	To avoid 
	misleading discrepancies between data and SUSY calculations  
	it is, however, preferable 
	to use a mixed spin-scalar coupling approach.
	This approach is applied to estimate 
	future prospects of experiments with the odd-neutron 
	high-spin isotope $^{73}$Ge.
	It is noticed that the DAMA evidence favors the 
	light Higgs sector in the effMSSM, 
	a high event rate in a $^{73}${Ge} detector and 
	relatively high upgoing muon fluxes 
	from relic neutralino annihilations 
	in the Earth and the Sun. 
\end{abstract} 

\maketitle 

\section{Introduction}
        Nowadays the main efforts and expectations 
	in the direct dark matter searches 
        are concentrated in the field of so-called  
        spin-independent (or scalar) interaction of a dark matter 
	Weakly Interacting Massive Particle (WIMP) with a target nucleus. 
        The lightest supersymmetric (SUSY) particle (LSP) neutralino is 
        assumed to be the best WIMP dark matter (DM) candidate.  
        It is believed that 
	for heavy enough nuclei
	this spin-independent (SI) interaction of 
        DM particles with nuclei usually gives the 
	dominant contribution to the expected event rate of its detection.
        The reason is the strong (proportional to the squared mass 
        of the target nucleus) 
        enhancement of SI WIMP-nucleus interaction.
        The results 
	currently obtained in the field are
        usually presented in the form of exclusion curves
	due to non-observation of the WIMPs.
        For a fixed mass of the WIMP the cross sections
        of SI elastic WIMP-nucleon interaction
        located above these curves are excluded.

\begin{figure}[t!] 
\begin{picture}(60,135) 
\put(-63,-63){\includegraphics{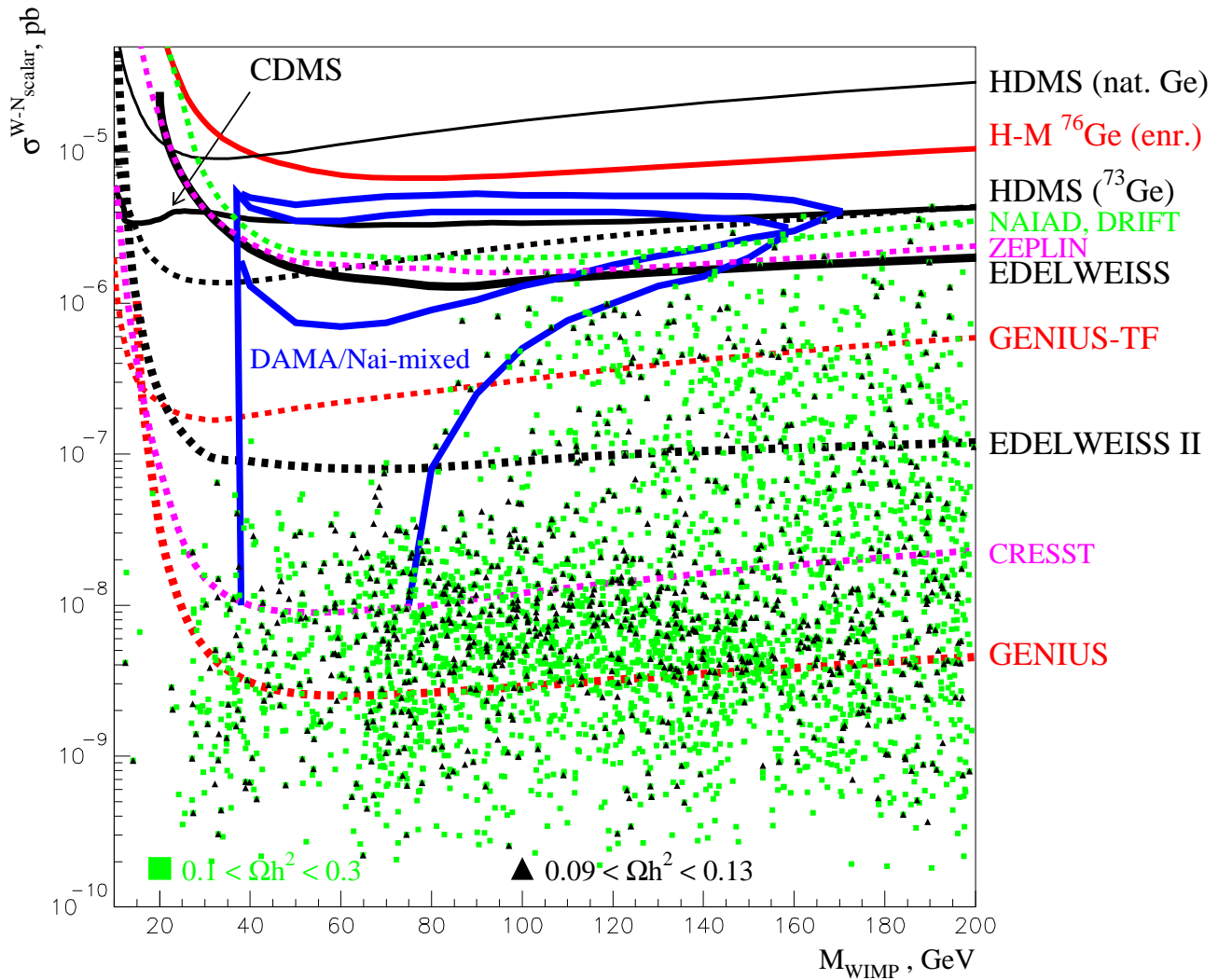}}
\end{picture}
\caption{WIMP-nucleon cross section 
        limits in pb for scalar (spin-independent) interactions as 
        a function of the WIMP mass in GeV. 
        Shown are contour lines for some of the present experimental limits 
        (solid lines) and some of projected experiments (dashed lines). 
        The closed DAMA/NaI contour corresponds to a complete neglection 
	of spin-dependent WIMP-nucleon interaction
	($\sigma^{}_{\rm SD}=0$), 
	while the open contour is obtained with the assumption that
	$\sigma^{}_{\rm SD}=0.08$ pb
\cite{Bernabei:2003za}. 
        Our theoretical expectations are shown by scatter plots
	for a relic neutralino density 
	$0.1 < \Omega_\chi h^2_0<0.3$ (green boxes) and to 
	WMAP relic density $0.094 < \Omega_\chi h^2_0<0.129$ 
	(black triangles).
	Similar estimations one can find for example in
\cite{Bednyakov:2000he,Ellis:2003ry,Ellis:2003eg}. 
	}
\label{Scalar-2003} 
\end{figure}

	Only the DAMA collaboration claims 
        observation of first evidence for the dark matter signal,
        due to registration of the annual modulation effect
\cite{Bernabei:2000qi,Bernabei:2003za,Bernabei:2003wy}.
	The DAMA results are shown in the middle of 
Fig.~\ref{Scalar-2003} as two 
	contours together with some set
	of other exclusion curves already obtained (solid lines)
	and expected in the future (dashed lines).
	Aimed since more than one decade at 
	the DM particle direct detection,  
	the DAMA experiment (DAMA/NaI) with 100 kg
	of highly radio-pure NaI(Tl) scintillator detectors  
	successfully operated till July 2002 
	at the Gran Sasso National Laboratory of the I.N.F.N.
	On the basis of the results obtained over 7 annual cycles 
	(107731~kg$\cdot$day total exposure)
	the presence of a WIMP model-independent 
	annual modulation signature was demonstrated 
	and the WIMP presence in the galactic halo is strongly supported 
	at 6.3 $\sigma$ C.L. 
\cite{Bernabei:2003za}.
	The main result of the DAMA observation 
	of the annual modulation signature 
	is the low-mass region of the WIMPs
	($40 < m_\chi < 150$~GeV), provided these WIMPs are
	cold dark matter particles. 
	No other experiment at present has the sensitivity 
	to look for this modulation effect.

	It is obvious that such a serious claim should be verified 
	at least by one other completely independent experiment.  
	To confirm this DAMA result one should perform
	a new experiment which would have (in reasonable time) 
	the same or better sensitivity to the annual modulation signal
	(and also it would be better to locate this new 
	setup in another low-background underground laboratory).
	This mission, in particular, 
	could be executed by new-generation experiments
	with large enough mass of germanium HP detectors both 
	with spin ($^{73}$Ge) and spin-less (natural Ge). 
        Due to kinematic reasons 
($M_{\rm Target}\approx M_{\rm WIMP}$)
	these germanium isotopes with their masses being almost 
	equal to the mass of the DAMA WIMP 
	(about 70 GeV) have the best efficiency for such WIMP
	detection.
	A new setup with ``naked'' Ge detectors in liquid nitrogen
	(GENIUS-TF) is already installed 
	and works over months under the low-background conditions
	of the Gran Sasso Laboratory 
\cite{Klapdor-Kleingrothaus:2003pg}.
	The GENIUS-TF experiment is planned 
	to be sensitive to the annual modulation signal
	with data taking over about 5 years 
	with a large enough mass of the Ge detectors 
\cite{Tomei:2003vc}. 

	In this paper we start from   
	the final results of the DAMA collaboration
	based on the 7-year-long measurements of the annual modulation 
\cite{Bernabei:2003za}
	and consider their possible consequences for 
	dark matter search with high-spin $^{73}$Ge detectors like HDMS
\cite{Klapdor-Kleingrothaus:2002pg}.
        We also briefly consider some aspects of the spin-dependent 
        (or axial-vector) interaction of the DM WIMPs with nuclei.
        There are at least three reasons to think that this 
        SD interaction could be very important. 
        First, contrary to the only one constraint for SUSY models available 
        from the scalar WIMP-nucleus interaction, the spin WIMP-nucleus 
        interaction supplies us with two such constraints (see for example 
\cite{Bednyakov:1994te} and formulas below).
        Second, one can notice 
\cite{Bednyakov:2000he,Bednyakov:2002mb}
        that even with a very sensitive DM detector
        (say, with a sensitivity of $10^{-5}\,$events$/$day$/$kg)
        which is sensitive only to the WIMP-nucleus 
        scalar interaction (with spin-less target nuclei) 
        one can, in principle, miss a DM signal. 
        To safely avoid such a situation one should
        have a spin-sensitive DM detector, i.e. a detector 
        with spin-non-zero target nuclei.
        Finally, there is a complicated 
        nuclear spin structure, which for example, possesses 
        the so-called long $q$-tail form-factor behavior. 
        Therefore for heavy mass target nuclei and heavy WIMP masses
	the SD efficiency to detect a DM signal 
        is much higher than the SI efficiency
\cite{Engel:1991wq}.  

\section{Approach to our calculations}  
\subsection{Cross sections and event rate}
        A dark matter event is elastic scattering 
        of a relic neutralino $\chi$ 
        from a target nucleus $A$ producing a nuclear 
        recoil $E_{\rm R}$ which can be detected by a suitable detector.
        The differential event rate in respect to the recoil 
        energy is the subject of experimental measurements.
        The rate depends on the distribution of
        the relic neutralinos in the solar vicinity $f(v)$ and
        the cross section of neutralino-nucleus elastic scattering
\cite{Bernabei:2003za,Jungman:1996df,Lewin:1996rx,Smith:1990kw,%
Bednyakov:1999yr,Bednyakov:1996yt,Bednyakov:1997ax,Bednyakov:1997jr,%
Bednyakov:1994qa}.
        The differential event rate per unit mass of 
        the target material has the form
\begin{equation}
\label{Definitions.diff.rate}
        \frac{dR}{dE_{\rm R}} = N_T \frac{\rho_\chi}{m_\chi}
        \displaystyle
        \int^{v_{\rm max}}_{v_{\rm min}} dv f(v) v
        {\frac{d\sigma}{dq^2}} (v, q^2). 
\end{equation}
        The nuclear recoil energy
        $E_{\rm R} = q^2 /(2 M_A )$ is typically about $10^{-6} m_{\chi}$ and 
        $N_T={\cal N}/A$ is the number density of target nuclei, where 
        ${\cal N}$ is the Avogadro number and $A$ is the atomic mass
        of the nuclei with mass $M_A$. 
        The neutralino-nucleus elastic scattering cross section 
        for spin-non-zero ($J\neq 0$) 
        nuclei contains SI and SD terms 
\cite{Engel:1992bf,Engel:1991wq,Ressell:1993qm}: 
\begin{eqnarray}
\nonumber
{\frac{d\sigma^{A}}{dq^2}}(v,q^2) 
        &=& \frac{\sum{|{\cal M}|^2}}{\pi\, v^2 (2J+1)} 
         =  \frac{S^A_{\rm SD} (q^2)}{v^2 (2J+1)} 
           +\frac{S^A_{\rm SI} (q^2)}{v^2 (2J+1)} \\
\label{Definitions.cross.section}
        &=& \frac{\sigma^A_{\rm SD}(0)}{4\mu_A^2 v^2}F^2_{\rm SD}(q^2)
           +\frac{\sigma^A_{\rm SI}(0)}{4\mu_A^2 v^2}F^2_{\rm SI}(q^2).
\end{eqnarray} 
        The normalized $(F^2_{\rm SD,SI}(0) = 1)$
	non-zero-momentum-transfer nuclear form-factors
\begin{equation}
\label{Definitions.form.factors}
F^2_{\rm SD,SI}(q^2) = \frac{S^{A}_{\rm SD,SI}(q^2)}{S^{A}_{\rm SD,SI}(0)},
\end{equation}
        are defined via nuclear structure functions
\cite{Engel:1992bf,Engel:1991wq,Ressell:1993qm} 
\begin{eqnarray}
\label{Definitions.scalar.structure.function}
S^{A}_{\rm SI}(q) 
        &=& 
        \sum_{L\, {\rm even}} 
        \vert\langle J \vert\vert {\cal C}_L(q) \vert\vert J \rangle \vert^2 
        \simeq  
        \vert\langle J \vert\vert {\cal C}_0(q) \vert\vert J \rangle \vert^2 ,
\\[3pt]
\label{Definitions.spin.structure.function}
S^A_{\rm SD}(q) 
        &=& 
        \sum_{L\, {\rm odd}} \big( 
        \vert\langle N \vert\vert {\cal T}^{el5}_L(q) 
        \vert\vert N \rangle\vert^2 + \vert\langle N \vert\vert 
        {\cal L}^5_L (q) \vert\vert N \rangle\vert^2\big). 
\end{eqnarray} 
        The transverse electric ${\cal T}^{el5}(q)$ 
        and longitudinal ${\cal L}^5(q)$ multipole projections of the
        axial vector current operator, 
	and the scalar function ${\cal C}_L(q)$ are given in the form
\begin{eqnarray*}
{\cal T}^{el5}_L(q) 
        &=& \frac{1}{\sqrt{2L+1}}\sum_i\frac{a^{}_0 +a^{}_1\tau^i_3}{2}
                 \Bigl[
                -\sqrt{L}   M_{L,L+1}(q\vec{r}_i)
                +\sqrt{L+1} M_{L,L-1}(q\vec{r}_i)
                 \Bigr], \\
 {\cal L}^5_L(q)
         &=& \frac{1}{\sqrt{2L+1}}\sum_i
                  \Bigl( \frac{a^{}_0}{2} +
                     \frac{a^{}_1 m^2_\pi \tau^i_3}{2(q^2+m_\pi^2)}
                  \Bigr)                 
                  \Bigl[
                 \sqrt{L+1} M_{L,L+1}(q\vec{r}_i)
                +\sqrt{L}   M_{L,L-1}(q\vec{r}_i)
                 \Bigr], \\[5pt]
{\cal C}_L(q) 
        &=& \sum_{i,\ {\rm nucleons}} c_0^{} j_L(qr_i)Y_L(\hat{r}_i), \  \  \ 
{\cal C}_0(q) = \sum_{i} c_0^{} j_0(qr_i)Y_0(\hat{r}_i),  
\end{eqnarray*} 
        where $a^{}_{0,1} = a^{}_n \pm a^{}_p$ 
	and $M_{L,L'}(q\vec{r}_i) = 
	j_{L'}(qr_i)[Y_{L'}(\hat{r}_i)\vec{\sigma}_i]^L$
\cite{Engel:1992bf,Engel:1991wq,Ressell:1993qm}.
        The nuclear SD and SI cross sections at $q=0$ in 
(\ref{Definitions.cross.section}) can be presented as follows
\begin{eqnarray}
\label{Definitions.scalar.zero.momentum}
\sigma^A_{\rm SI}(0) 
        &=& \frac{4\mu_A^2 \ S^{}_{\rm SI}(0)}{(2J+1)}\! =\!
             \frac{\mu_A^2}{\mu^2_p}A^2 \sigma^{p}_{{\rm SI}}(0), \\ 
\label{Definitions.spin.zero.momentum}
\sigma^A_{\rm SD}(0)
        &=&  \frac{4\mu_A^2 S^{}_{\rm SD}(0)}{(2J+1)}\! =\!
             \frac{4\mu_A^2}{\pi}\frac{(J+1)}{J}
             \left\{a_p\langle {\bf S}^A_p\rangle 
                  + a_n\langle {\bf S}^A_n\rangle\right\}^2\\
\label{Definitions.spin.zero.momentum.Tovei}
      &=&
        \frac{\mu_A^2}{\mu^2_{p}}\frac{(J+1)}{3\, J}
\left\{ \sqrt{\sigma^{p}_{{\rm SD}}(0)}\langle{\bf S}^A_p\rangle 
       +{\rm sign}(a_p\, a_n)
        \sqrt{\sigma^{n}_{{\rm SD}}(0)}\langle{\bf S}^A_n\rangle\right\}^2
\\
&=&
	\frac{\mu_A^2}{\mu_p^2}
	\frac43 \frac{J+1}{J}
	\sigma^{pn}_{\rm SD}(0)	
	\left\{ \langle {\bf S}^A_p\rangle \cos\theta
               +\langle {\bf S}^A_n\rangle \sin\theta   
       \right\}^2.
\label{Definitions.spin.zero.momentum.Bernabei}
\end{eqnarray} 
        Here $\displaystyle \mu_A = \frac{m_\chi M_A}{m_\chi+ M_A}$
        is the reduced $\chi$-$A$ mass, 
	and $\mu_p=\mu_n$ is assumed.
	Following Bernabei et al.
\cite{Bernabei:2003za,Bernabei:2001ve}
	the effective spin WIMP-nucleon cross section
	$\sigma^{pn}_{\rm SD}(0)$
	and the coupling mixing angle $\theta$ were introduced
\begin{eqnarray}
\label{effectiveSD-cs}
\sigma^{pn}_{\rm SD}(0)
	&=& \frac{\mu_p^2}{\pi}\frac43 
		\Bigl[ a_p^2 +a_n^2 \Bigr], \qquad
\tan\theta = \frac{{a}_{n}}{{a}_{p}}; \\
\label{effectiveSD-cs-pn}
\sigma^p_{\rm SD}&=&\sigma^{pn}_{\rm SD} \cdot \cos^2 \theta, \quad
\sigma^n_{\rm SD}=\sigma^{pn}_{\rm SD} \cdot \sin^2 \theta.
\end{eqnarray}
        The zero-momentum-transfer proton and neutron SI and SD cross sections
\begin{eqnarray}
\label{Definitions.scalar.zero.cs}
\sigma^{p}_{{\rm SI}}(0) 
        = 4 \frac{\mu_p^2}{\pi}c_{0}^2,
&\qquad&
        c^{}_{0} \equiv c^{(p,n)}_0 = \sum_q {\cal C}_{q} f^{(p,n)}_q; \\
\label{Definitions.spin.zero.cs}
\sigma^{p,n}_{{\rm SD}}(0)  
        =  12 \frac{\mu_{p,n}^2}{\pi}{a}^2_{p,n} 
&\qquad&
        a_p =\sum_q {\cal A}_{q} \Delta^{(p)}_q, \quad 
        a_n =\sum_q {\cal A}_{q} \Delta^{(n)}_q
\end{eqnarray}
        depend on the effective neutralino-quark scalar 
        ${\cal C}_{q}$ and axial-vector ${\cal A}_{q}$ couplings 
        from the effective Lagrangian
\begin{equation}
\label{Definitions.effective.lagrangian}
{\cal  L}_{\rm eff} = \sum_{q}^{}\left( 
        {\cal A}_{q}\cdot
      \bar\chi\gamma_\mu\gamma_5\chi\cdot
                \bar q\gamma^\mu\gamma_5 q + 
        {\cal C}_{q}\cdot\bar\chi\chi\cdot\bar q q
        \right)     \ + ... 
\end{equation}
        and on the spin ($\Delta^{(p,n)}_q$)
        and mass ($f^{(p,n)}_q$) structure of nucleons.
	The parameters $a_{p(n)}$ in 
(\ref{Definitions.spin.zero.cs}) can be considered as effective
	WIMP-proton(neutron) couplings. 
        The factors $\Delta_{q}^{(p,n)}$ in 
(\ref{Definitions.spin.zero.cs}) parameterize the quark 
        spin content of the nucleon 
        and are defined by the relation
        $ \displaystyle 2 \Delta_q^{(n,p)} s^\mu  \equiv 
          \langle p,s| \bar{\psi}_q\gamma^\mu \gamma_5 \psi_q    
          |p,s \rangle_{(p,n)}$.
        A global QCD analysis for the $g_1$ structure functions
\cite{Mallot:1999qb} including ${\cal O}(\alpha_s^3)$ corrections
        supplied us with the values
\cite{Ellis:2000ds}
\begin{equation}
\label{Spin-update}
\Delta_{u}^{(p)} = \Delta_{d}^{(n)} =   0.78 \pm 0.02,  \quad 
\Delta_{d}^{(p)} = \Delta_{u}^{(n)} =  -0.48 \pm 0.02,  \quad 
\Delta_{s}^{(p)} = \Delta_{s}^{(n)} =  -0.15 \pm 0.02.
\end{equation}
	The nuclear spin (proton, neutron) operator is defined as follows 
\begin{equation}
\label{Definitions.spin.operator}
 {\bf S}_{p,n} = \sum_i^A {\bf s}_{p,n} ({i}),
\end{equation}
        where $i$ runs over all nucleons.
        Further the convention is used 
        that all angular momentum operators 
        are evaluated in their $z$-projection 
        in the maximal $M_J$ state, e.g.
\begin{equation}
\label{Definitions.spin.operator.1}
\langle {\bf S} \rangle \equiv \langle N \vert {\bf S} \vert N \rangle
\equiv  \langle J,M_J = J \vert S_z \vert J,M_J = J \rangle.
\end{equation}
        Therefore
        $ \langle {\bf S}_{p(n)} \rangle $ is the spin of the proton 
        (neutron) averaged over all nucleons in the nucleus $A$.
        The cross sections at zero momentum transfer show strong 
        dependence on the nuclear structure of the ground state
\cite{Engel:1995gw,Ressell:1997kx,Divari:2000dc}. 

        The relic neutralinos in the halo of our Galaxy have a mean velocity of
        $\langle v \rangle  \simeq 300~{\rm km/s} = 10^{-3} c$.  
        When the product $q_{\max}R \ll 1$, 
        where $R$ is the nuclear radius 
        and $q_{\max} = 2 \mu_A v$ is the maximum momentum 
        transfer in the $\chi$-$A$ scattering, the matrix element 
        for the SD 
	$\chi$-$A$ scattering reduces to a very simple form
({\em zero momentum transfer limit})\
\cite{Engel:1995gw,Ressell:1997kx}:
\begin{equation}
\label{Definitions.matrix.element}
 {\cal M} = C \langle N\vert a_p {\bf S}_p + a_n {\bf S}_n
        \vert N \rangle \cdot {\bf s}_{\chi}
          = C \Lambda \langle N\vert {\bf J}
         \vert N \rangle \cdot {\bf s}_{\chi}.
\end{equation}
        Here ${\bf s}_{\chi}$ is the spin of the neutralino, and 
\begin{equation}
 \Lambda = {{\langle N\vert a_p {\bf S}_p + a_n {\bf S}_n
\vert N \rangle}\over{\langle N\vert {\bf J}
\vert N \rangle}} =
{{\langle N\vert ( a_p {\bf S}_p + a_n {\bf S}_n ) \cdot {\bf J}
\vert N \rangle}\over{ J(J+1)
}}. 
\end{equation}
        It is seen that the $\chi$ couples to the spin carried
        by the protons and the neutrons.  
        The normalization $C$ involves the coupling
        constants, masses of the exchanged bosons and various LSP
        mixing parameters that have no effect upon the nuclear matrix element
\cite{Griest:1988ma}.  
        In the $q=0$  limit 
        the spin structure function 
(\ref{Definitions.spin.structure.function}) reduces to
\begin{equation}
S^A_{\rm SD}(0) = 
        {2 J + 1\over{\pi}} \Lambda^2 J(J + 1). 
\end{equation}

        The first model to estimate the spin content in the nucleus
        for the dark matter search was the 
        independent single-particle shell model ({ISPSM})
        used originally by Goodman and Witten 
\cite{Goodman:1985dc} and later in
\cite{Drukier:1986tm,Ellis:1988sh,Smith:1990kw}.
        Here the ground state value of the nuclear total spin $J$ can be 
        described by that of one extra nucleon interacting with the
        effective potential of the nuclear core.  
        There are nuclear structure calculations 
        (including non-zero-momentum approximation) for spin-dependent
        neutralino interaction with 
        helium $^3$He 
\cite{Vergados:1996hs};
        fluorine $^{19}$F
\cite{Vergados:2002bb,Divari:2000dc,Vergados:1996hs};
        sodium $^{23}$Na
\cite{Vergados:2002bb,Ressell:1997kx,Divari:2000dc,Vergados:1996hs};
        aluminum $^{27}$Al
\cite{Engel:1995gw};
        silicon $^{29}$Si
\cite{Vergados:2002bb,Ressell:1993qm,Divari:2000dc};
        chlorine $^{35}$Cl
\cite{Ressell:1993qm};
        potassium $^{39}$K
\cite{Engel:1995gw};
        germanium $^{73}$Ge 
\cite{Ressell:1993qm,Dimitrov:1995gc};
        niobium $^{93}$Nd
\cite{Engel:1992qb};
        iodine $^{127}$I
\cite{Ressell:1997kx};
        xenon $^{129}$Xe
\cite{Ressell:1997kx} and
        $^{131}$Xe
\cite{Ressell:1997kx,Nikolaev:1993vw,Engel:1991wq}; 
        tellurium $^{123}$Te
\cite{Nikolaev:1993vw}
        and  $^{125}$Te
\cite{Ressell:1997kx};
        lead $^{208}$Pb
\cite{Kosmas:1997jm,Vergados:1996hs}.
        The zero-momentum case is also investigated for 
        Cd, Cs, Ba and La in 
\cite{Pacheco:1989jz,Nikolaev:1993vw,Iachello:1991ut}.

        There are several approaches 
	to more accurate calculations of the nuclear
        structure effects relevant to the dark matter detection.
	The list of the models includes the
        Odd Group Model ({OGM}) of Engel and Vogel
\cite{Engel:1989ix} and their extended OGM ({EOGM})
\cite{Engel:1989ix,Engel:1992bf}; 
        Interacting Boson Fermion Model ({IBFM}) of
        Iachello, Krauss, and Maino      
\cite{Iachello:1991ut};
        Theory of Finite Fermi Systems ({TFFS}) of 
        Nikolaev and Klapdor-Kleingrothaus
\cite{Nikolaev:1993dd};
        Quasi Tamm-Dancoff Approximation ({QTDA}) of Engel
\cite{Engel:1991wq};
        different shell model treatments ({SM}) by Pacheco and Strottman 
\cite{Pacheco:1989jz};
        by Engel, Pittel, Ormand and Vogel 
\cite{Engel:1992qb} and 
        Engel, Ressell, Towner and Ormand,
\cite{Engel:1995gw},    
        by Ressell et al.
\cite{Ressell:1993qm} and 
        Ressell and Dean
\cite{Ressell:1997kx};
        by Kosmas, Vergados et al.
\cite{Vergados:1996hs,Kosmas:1997jm,Divari:2000dc};
        the so-called ``{hybrid}'' model of Dimitrov, Engel and Pittel 
\cite{Dimitrov:1995gc}
        and perturbation theory 
	based on calculations of Engel et al.
\cite{Engel:1995gw}.

	The direct detection rate 
(\ref{Definitions.diff.rate}) in 
	a nucleus $A$ integrated over 
	the recoil energy interval from threshold energy, $\eth$, 
	till maximal energy, $\emx$, is a sum of SD and SI contributions: 
\begin{eqnarray}
\label{for-toy-mixig}
R(\eth, \emx)&=&
	\alpha(\eth,\emx,m_\chi)\,\sigma^p_\SI
	+\beta(\eth,\emx,m_\chi)\,\sigma^{pn}_\SD;\\
&&\alpha(\eth,\emx,m_\chi)
	= N_T \frac{\rho_\chi M_A}
		{2 m_\chi \mu_p^2 } A^2 
	A_\SI(\eth,\emx),\nonumber \\
&&\beta(\eth,\emx,m_\chi)
 	=	N_T \frac{\rho_\chi M_A}
		 {2 m_\chi \mu_p^2 } 
	\frac43 \frac{J+1}{J}
	\left( \langle {\bf S}^A_p\rangle \cos\theta
              +\langle {\bf S}^A_n\rangle \sin\theta   
       \right)^2
	A_\SD(\eth,\emx); \nonumber \\
&& A_{\SI,\SD}(\eth,\emx) = 
	\frac{\langle {v}\rangle}
	 {\langle {v}^2 \rangle}
	\int_\eth^\emx d\ER F^2_{\SI,\SD}(\ER)I(\ER).
\label{for-toy-mixig-FF}
\end{eqnarray}
	To estimate the event rate 
(\ref{for-toy-mixig})	
	one needs to know a number of quite uncertain 
	astrophysical and nuclear structure parameters
	as well as the precise characteristics of the experimental setup
	(see, for example, the discussions in 
\cite{Bernabei:2003xg,Bernabei:2003za}).

\subsection{Effective low-energy MSSM}  
	To obtain as much as general predictions
	it appeared  more convenient to work within 
	a phenomenological SUSY model 
	whose parameters are defined 
	directly at the electroweak scale, 
	relaxing completely constraints following from 
	any unification assumption
	as for example in  
\cite{Bednyakov:1999vh,Mandic:2000jz,Bergstrom:1996cz,%
Gondolo:2000fh,Bednyakov:1998is,Bergstrom:2000pn}, 
	and which is called an effective 
	scheme of MSSM (effMSSM) in 
\cite{Bottino:2000jx}, and later by some people 
	low-energy effective supersymmetric theory (LEEST) in 
\cite{Ellis:2003ry,Ellis:2003eg}. 
	In our previous calculations in effMSSM
\cite{Bednyakov:2003wf,Bednyakov:2002dz,Bednyakov:2002js,%
Bednyakov:2002mb,Bednyakov:2000he,Bednyakov:2002ng,%
Bednyakov:1999vh,Bednyakov:1997jr,%
Bednyakov:1997ax,Bednyakov:1994qa} we have adopted some effective
	scheme (with non-universal scalar masses and with 
	non-universal gaugino soft masses) which lead to large values for  
	direct detection rates of DM neutralinos.

	Our MSSM parameter space is determined by the entries of the mass 
	matrices of neutralinos, charginos, Higgs bosons, 
	sleptons and squarks. 
	The relevant definitions one can find in 
\cite{Bednyakov:1999vh}.
	The list of free parameters includes: 
	$\tan\beta$ is the ratio
	of neutral Higgs boson vacuum expectation values, 
	$\mu$ is the bilinear Higgs parameter of the superpotential,
	$M_{1,2}$ are soft gaugino masses, 
	$M_A$ is the CP-odd Higgs mass, 
	$m^2_{\widetilde Q}$, $m^2_{\widetilde U}$, $m^2_{\widetilde D}$ 
	($m^2_{\widetilde L}$, $m^2_{\widetilde E}$) are 
	squared squark (slepton) 
	mass parameters for the 1st and 2nd generation,        
	$m^2_{\widetilde Q_3}$, $m^2_{\widetilde T}$, $m^2_{\widetilde B}$ 
	($m^2_{\widetilde L_3}$, $m^2_{\widetilde \tau}$) 	
	are squared squark (slepton) mass parameters 
	for 3rd generation 
	and $A_t$, $A_b$, $A_\tau$ are soft trilinear 
	couplings for the 3rd generation.
	The third gaugino mass parameter $M_3$ defines the 
	mass of the gluino in the model and is 
	determined by means of the GUT assumption $M_2 = 0.3\, M_3$.

	Contrary to our previous considerations
\cite{Bednyakov:2003wf,Bednyakov:2002dz,Bednyakov:2002js,%
Bednyakov:2002mb,Bednyakov:2000he,Bednyakov:2002ng,%
Bednyakov:1999vh,Bednyakov:1997jr,%
Bednyakov:1997ax,Bednyakov:1994qa} 
	and aiming at exploration of the MSSM parameter space
	in the DAMA-inspired  
	domain of the lower masses of the LSP ($m_\chi < 200$~GeV), 
	we narrowed in the present work 
	the intervals of the randomly scanned parameter space to the 
	following:
\begin{eqnarray}
\nonumber
&-200{\rm ~GeV} < M_1 < 200{\rm ~GeV}, \quad
-1{\rm ~TeV} < M_2, \mu < 1{\rm ~TeV}, \quad
-2{\rm ~TeV} < A_t < 2{\rm ~TeV},&
\\ \label{Scanning}
&10 < \tan\beta < 50, \quad
50{\rm ~GeV} < M_A < 500{\rm ~GeV},&\\
\nonumber
&10{\rm ~GeV}^2 < m^2_{\widetilde{Q},\widetilde{Q}_3}, 
m^2_{\widetilde{L},\widetilde {L}_3} <10^6{\rm ~GeV}^2.&
\end{eqnarray}
	As previously we assume that squark masses are basically degenerate.
	Bounds on flavor-changing neutral currents 
	imply that squarks with equal gauge
	quantum numbers must be close in mass
\cite{Drees:2000he,Arnowitt:2000hi,Corsetti:2000xm}. 
	With the possible exception of third generation squarks
	the assumed degeneracy 
	holds almost model-independently
\cite{Drees:2000he}. 
	Therefore for other sfermion mass parameters as before in 
\cite{Bednyakov:2003wf,Bednyakov:2002dz,Bednyakov:2002js,%
Bednyakov:2002mb,Bednyakov:2000he,Bednyakov:2002ng,Bednyakov:1999vh} 
	we used the relations 
	$m^2_{\widetilde U_{}} = m^2_{\widetilde D_{}} 
	= m^2_{\widetilde Q_{}}$, 
	$m^2_{\widetilde E_{}} = m^2_{\widetilde {L}}$, 
	$m^2_{\widetilde T} = m^2_{\widetilde B} = m^2_{\widetilde{Q}_3}$,  
	$m^2_{\widetilde{E}_{3}} = m^2_{\widetilde{L}_3}$.
	The parameters $A_b$ and $A_\tau$ are fixed to be zero.

	We have included the current experimental 
	upper limits on sparticle and Higgs masses
	from the Particle Data Group 
\cite{Hagiwara:2002fs}. 
	For example, we use as previously 
	the following lower bounds for the SUSY particles:
$M_{\tilde\chi^\pm_{1,2}} \ge 100\,$GeV for charginos, 
$ M_{\tilde{\chi}^0_{1,2,3}} \geq 45, 76, 127$~GeV for non-LSP
					neutralinos, respectively;
$ M_{\tilde{\nu}}      \geq 43$~GeV for sneutrinos,
$ M_{\tilde{e}_R}      \geq 70$~GeV for selectrons,
$ M_{\tilde{q}}       \geq 210$~GeV  for squarks,
$ M_{\tilde{t}_1}      \geq 85$~GeV  for light top-squark,   
$ M_{H^0}              \geq 100$~GeV  for neutral Higgs bosons,
$ M_{\rm H^+}           \geq 70$~GeV  for the charged Higgs boson.
	Also the limits on the rare $b\rightarrow s \gamma$ decay 
\cite{Alam:1995aw,Abe:2001fi} following 
\cite{Bertolini:1991if,Barbieri:1993av,Buras:1994xp,Ali:1993ct} 
	have been imposed. 

	For each point in the MSSM parameter space (MSSM model) 
	we have evaluated the relic density of the light neutralinos
	$\Omega_{\chi} h^2_0$ with our code 
\cite{Bednyakov:2002dz,Bednyakov:2002js,Bednyakov:2002ng} based on 
\cite{Gondolo:2000ee}, 
	taking into account all coannihilation channels with 
	two-body final states that can occur between neutralinos, charginos,
	sleptons, stops and sbottoms,
	as long as their masses are $m_i<2m_\chi$.
	We assume as before 
	$0.1< \Omega_\chi h^2  < 0.3$ 
	for the cosmologically interesting region and
	we also consider the WMAP reduction of the region
	to $0.094< \Omega_\chi h^2  < 0.129$ 
\cite{Spergel:2003cb,Bennett:2003bz}
	and a possibility the LSP to be 
	not a unique DM candidate 
	with much smaller relic density 
	$0.002< \Omega h^2  < 0.1$. 

\section{Results and Discussions}
\subsection{Cross sections in the effMSSM for \boldmath $m_\chi < 200$~GeV}
\begin{figure}[!h] 
\begin{picture}(100,140)
\put(-30,-5){\includegraphics{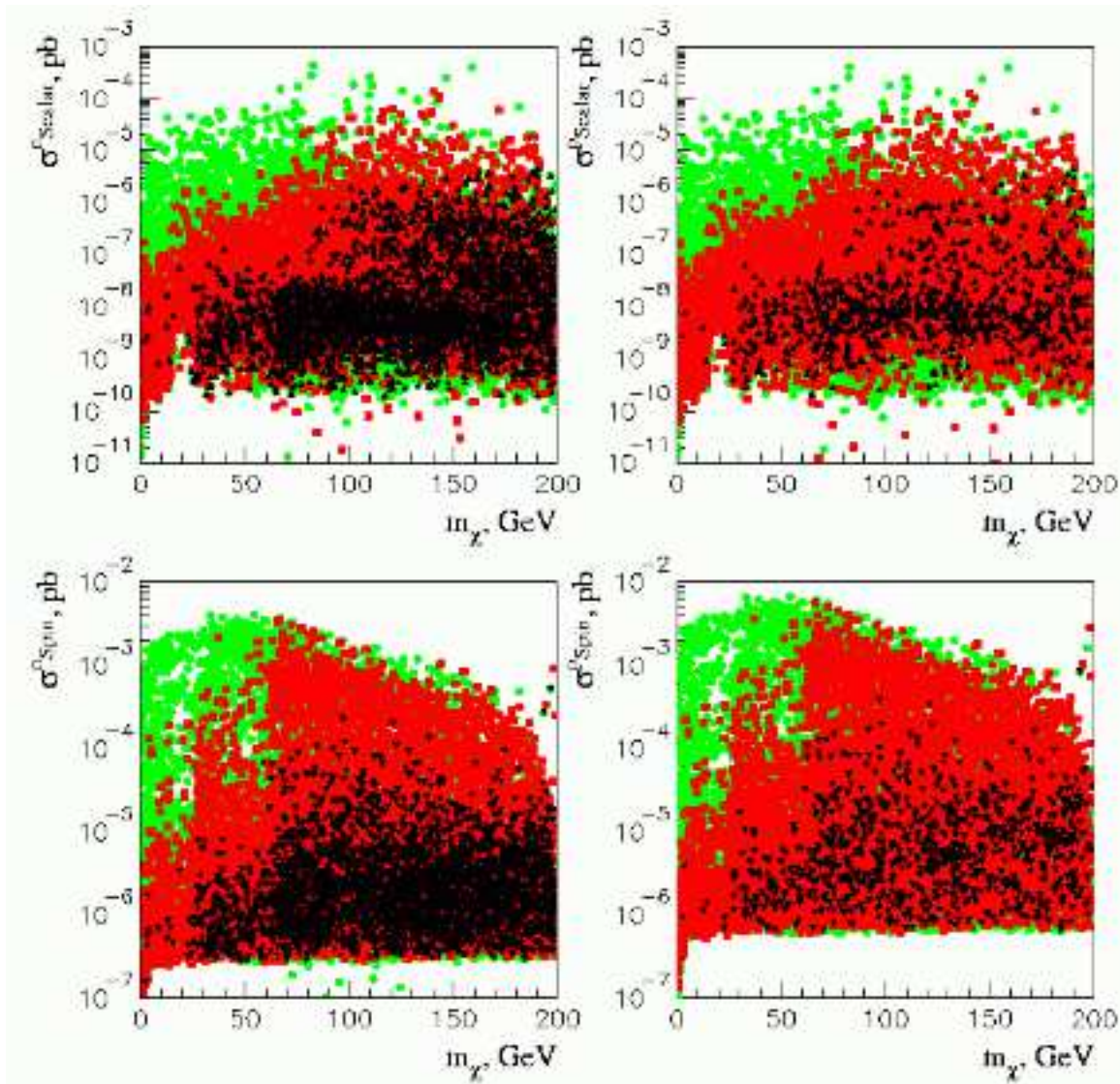}}
\end{picture}
\caption{Cross sections of the spin-dependent (spin) and 
	the spin-independent (scalar)
	interactions of WIMPs with the proton and the neutron.
	Filled green circles correspond to the relic neutralino density 
	$0< \Omega_\chi h^2_0<1$,
	red squares correspond to the sub-dominant relic neutralino
	contribution $0.002 < \Omega_\chi h^2_0<0.1$ 	
	and black triangles correspond to the relic neutralino density 
	$0.1 < \Omega_\chi h^2_0<0.3$ (left panel) and to 
	the WMAP relic density
	$0.094 < \Omega_\chi h^2_0<0.129$ (right panels). 
\label{CrossSections-vs-lsp}}
\end{figure}

\begin{figure}[t!] 
\begin{picture}(100,140)
\put(-25,-4){y\includegraphics{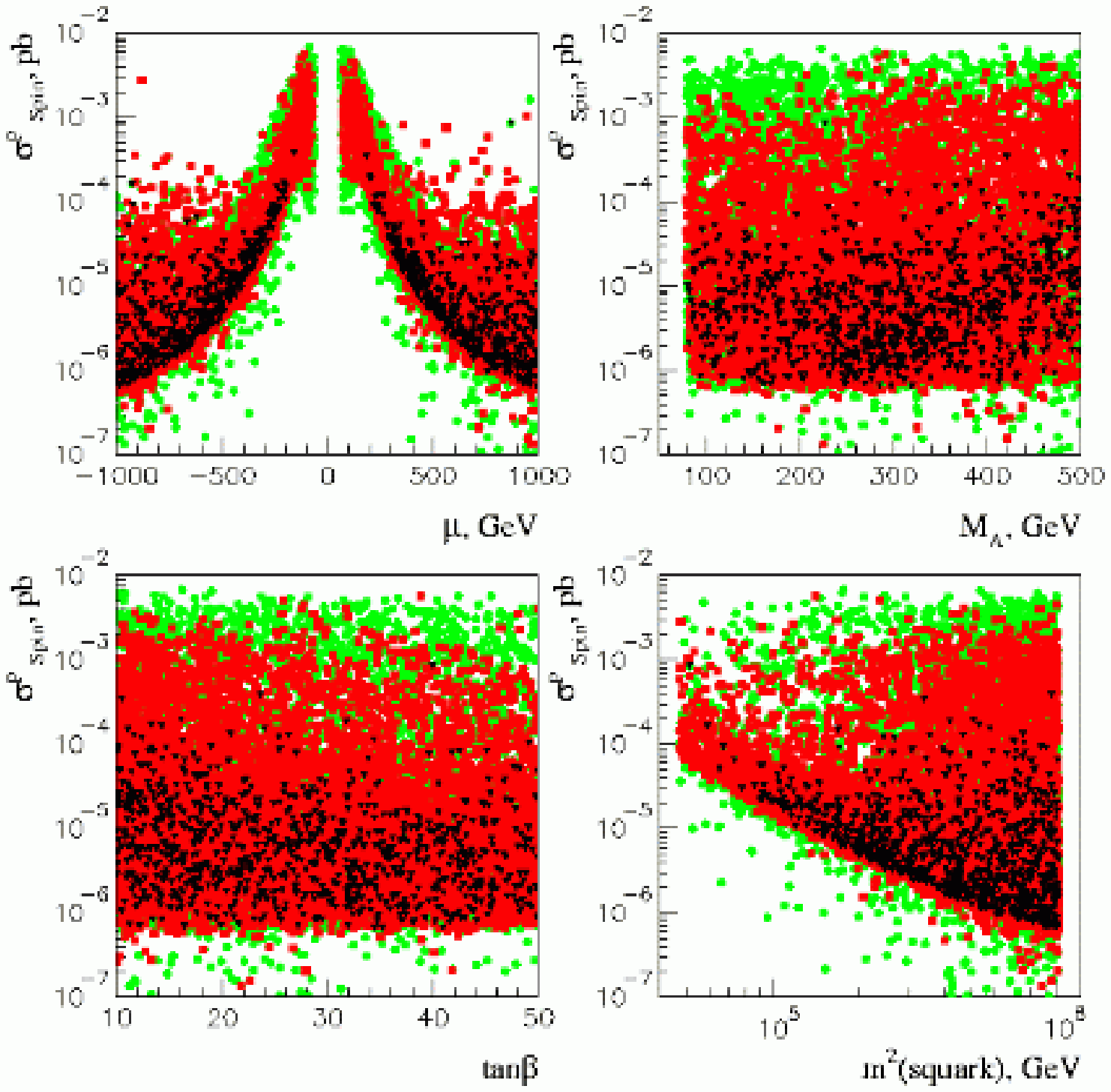}}
\end{picture}
\caption{Cross sections of of WIMP-proton spin-dependent
	interactions as function of 
	input parameters $\mu$, $M_A$, $\tan\beta$
	and $m^2_{\widetilde{Q}}$ with the same notations as in 
Fig.~\ref{CrossSections-vs-lsp} 
	for used constraints on the neutralino relic density
	$\Omega_\chi h^2$.
\label{CrossSections-vs-mssm-spin}}
\end{figure}

        The results of our evaluations of the
	zero-momentum-transfer proton and neutron SI 
(\ref{Definitions.scalar.zero.cs}) and SD 
(\ref{Definitions.spin.zero.cs}) 
	cross sections in the effMSSM approach within the  
	DAMA-inspired parameter space of 
(\ref{Scanning}) are shown as scatter plots 
in Figs.~\ref{CrossSections-vs-lsp}--%
\ref{CrossSections-vs-mssm-spin}.

	Scatter plots with individual cross sections of SD 
	and SI interactions of WIMPs with proton and neutron are given in 
Fig.~\ref{CrossSections-vs-lsp} as functions of the LSP mass.
	In the figure filled green circles correspond to cross sections
	calculated when the neutralino relic density 
	should just not overclose the Universe 
	($0.0<\Omega_\chi h^2_0<1.0$).
	Filled red squares show the same cross sections 
	when one assumes the relic neutralinos to be not the only 
	DM particles and give only sub-dominant 
	contribution to the relic density  
 	$0.002 < \Omega_\chi h^2_0<0.1$.
 	In the left panel of 
Fig.~\ref{CrossSections-vs-lsp}
	these cross sections are shown with the  black triangles
	corresponding to the case when the relic neutralino density 
	is in the bounds previously associated with a so-called 
	flat and accelerating Universe
	$0.1 < \Omega_\chi h^2_0<0.3$.
	The black triangles in the right panel in
Fig.~\ref{CrossSections-vs-lsp} 
	correspond to imposing the new WMAP
\cite{Spergel:2003cb,Bennett:2003bz}
	constraint on matter relic density   
	$0.094 < \Omega_\chi h^2_0<0.129$. 
	Despite a visible reduction   
	of the allowed domain for the relic density due to the WMAP result 
	the upper bounds for the 
	spin-dependent and the spin-independent WIMP-nucleon cross section 
	are not significantly affected.
	From the comparison of circle and square distributions,
	as expected, follows that the largest cross section values
	correspond to smallest values of the $\Omega_\chi$,
	especially for smaller LSP masses. 
	It is seen that the LSP as a sub-dominant DM particle
	favors the large SD and SI cross sections. 
	Furthermore the maximal SD and SI cross sections in
Fig.~\ref{CrossSections-vs-lsp} (green circles)
	come for very small relic density values
 	$0.0< \Omega_\chi h^2_0<0.002$.
\begin{figure}[h!] 
\begin{picture}(100,140)
\put(-25,-4){\includegraphics{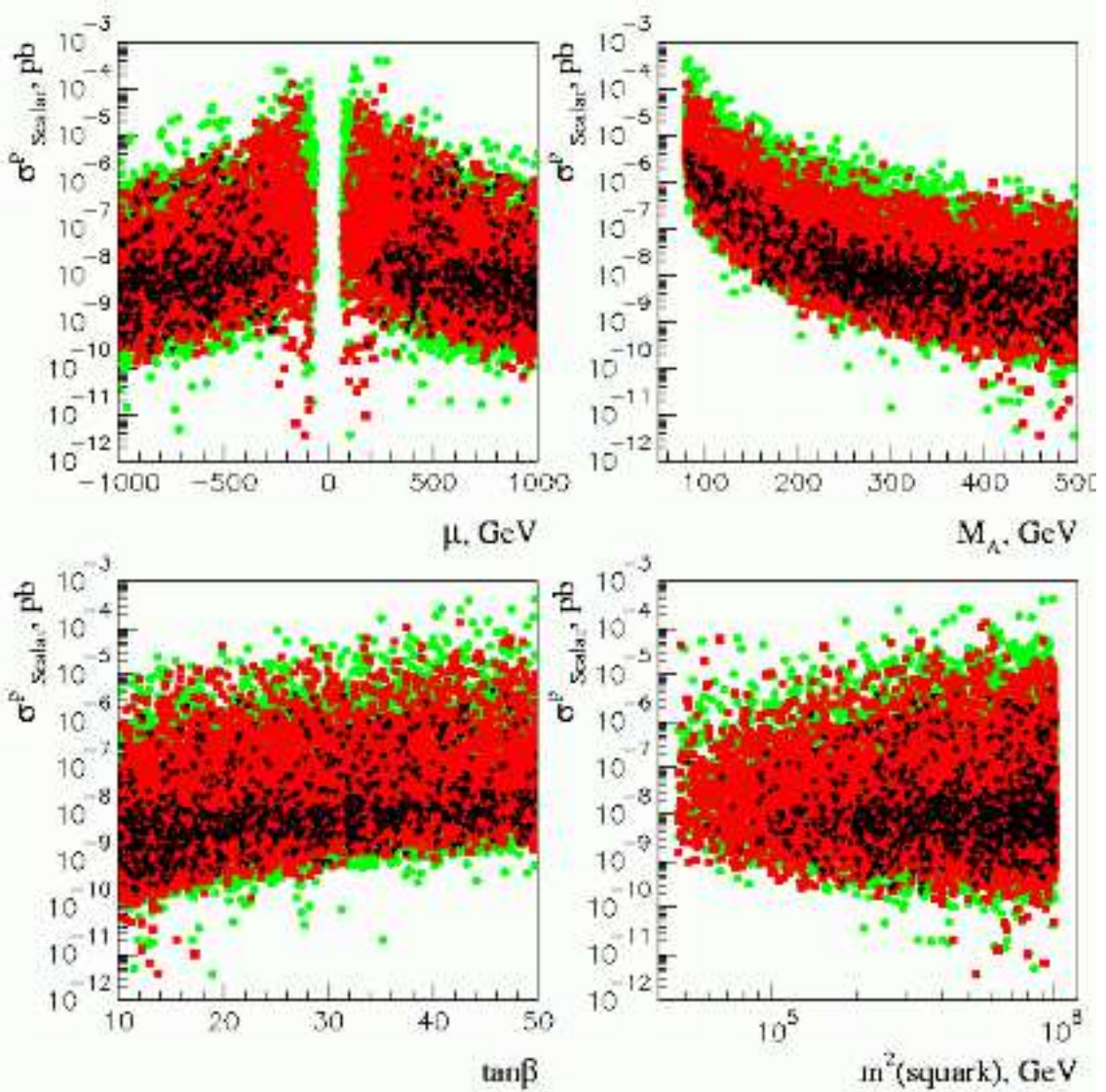}}
\end{picture}
\caption{Cross sections of of WIMP-proton spin-independent
	interactions as function of 
	input parameters $\mu$, $M_A$, $\tan\beta$
	and $m^2_{\widetilde{Q}}$ with the same notations as in 
Fig.~\ref{CrossSections-vs-lsp} 
	for used constraints on the neutralino relic density
	$\Omega_\chi h^2$.
\label{CrossSections-vs-mssm-scalar}}
\end{figure} 

	One can see also that in our effMSSM with 
	parameters from 
(\ref{Scanning})
	the lower bound value in the 
	relic density constraints (as for example,
	$0.094$ in the case of WMAP) 
	restricts from below the allowed masses of the LSP
	in accordance with previous considerations
\cite{Bednyakov:1997ax,Bottino:2003cz}.
		
	The spin-dependent and spin-independent WIMP-proton
	cross sections as functions of input MSSM parameters
	$\mu$, $M_A$, $\tan\beta$ and $m^2_{\widetilde{Q}}$ are shown in 
Figs.~\ref{CrossSections-vs-mssm-scalar} and
\ref{CrossSections-vs-mssm-spin}. 
	There is no noticeable dependence of these scatter plots 
	on the other free parameters from our set 	
(\ref{Scanning}).
	From these figures one can see the similarity of the scatter plots  
	for spin-dependent and and scalar cross sections as functions of  
	$\mu$ and $m^2_{\widetilde{Q}}$. 
	Decrease of both lower bounds of the cross sections 
	with $m^2_{\widetilde{Q}}$ occur due to increase
	of masses of squarks, which enter the s-channel intermediate states.
	Both spin-dependent and spin-independent
	cross sections increase when $|\mu|$ decreases, 
	in agreement with literature 
\cite{Feng:2000gh,Mandic:2000jz,Accomando:1999eg}
	and our previous calculations
\cite{Bednyakov:1999vh,Bednyakov:2000he}.
	The increase of the scalar cross sections generally is connected 
	with an increase of the Higgsino admixture of the LSP and increase 
	of Higgsino-gaugino interference which enters this cross section
\cite{Feng:2000gh,Gabrielli:2000uy,Accomando:1999eg}.
	The reason of the Higgsino growth can be 
	non-universality of scalar soft masses
\cite{Accomando:1999eg}, variation of intermediate unification scale
\cite{Gabrielli:2000uy}, or focus point regime of the supersymmetry 
\cite{Feng:2000gh}.
	There is no any visible sensitivity of the SD cross sections
	to $\tan\beta$ and $M_A$ (Higgs bosons do not contribute)
	but the SI cross section possesses
	remarkable dependence on these parameters.
 	The SI cross sections rather quickly drop with growth
	of the CP-odd Higgs mass $M_A$ and increase with 
	$\tan\beta$ 
\cite{Bednyakov:1994qa,Baer:1998ai,Bednyakov:1999vh,%
Accomando:1999eg,Gabrielli:2000uy,Lahanas:2000xd,Bottino:2000jx}.
	The different $\tan\beta$- and $M_A$-dependence 
	of the SD and SI cross section
	as well as the general about-2-order-of-magnitude 
	excess of the spin-dependent cross sections over spin-independent
	cross sections may be important for observations
\cite{Tovey:2000mm,Ullio:2000bv,Bednyakov:2000he,%
Bednyakov:2003wf,Bednyakov:2002mb}.
	It is interesting to note, that maximal values
	for the LSP-proton SD cross section one can obtain
	in the pure Higgsino case (when only $Z$-exchange contributes) 
	at a level of ${5 \cdot 10^{-2} {\rm pb}}$.
	This value is almost reached by points from our scatter plots in
Figs.~\ref{CrossSections-vs-lsp} and
\ref{CrossSections-vs-mssm-spin}.

\subsection{Constraints on WIMP-nucleon spin interactions}
        For the spin-zero nuclear target the experimentally 
	measured event rate 
(\ref{Definitions.diff.rate}) 
	of direct DM particle detection, via formula 
(\ref{Definitions.cross.section}) 
	is connected with the zero-momentum WIMP-proton(neutron)
        cross section 
(\ref{Definitions.scalar.zero.momentum}). 
        The zero momentum scalar WIMP-proton(neutron) cross section 
        $\sigma^{p}_{{\rm SI}}(0)$ can be expressed through 
        effective neutralino-quark couplings ${\cal C}_{q}$
(\ref{Definitions.effective.lagrangian}) by means of expression 
(\ref{Definitions.scalar.zero.cs}).
        These couplings ${\cal C}_{q}$ (as well as ${\cal A}_{q}$) 
        can be directly connected with the
        fundamental parameters of a SUSY model 
        such as $\tan \beta$, $M_{1,2}$, $\mu$, masses 
        of sfermions and Higgs bosons, etc. 
        Therefore experimental limitations on the 
        spin-independent neutralino-nucleon cross section
        supply us with a constraint on the fundamental parameters
        of an underlying SUSY model.

        In the case of the spin-dependent WIMP-nucleus interaction
        from a measured differential rate 
(\ref{Definitions.diff.rate}) one first extracts 
        a limitation for $\sigma^{A}_{{\rm SD}}(0)$, 
        and therefore has in principle two constraints
\cite{Bednyakov:1994te}
        for the neutralino-proton $a^{}_p$ and  
        neutralino-neutron $a^{}_n$ effective spin couplings, 
	as follows from relation 
(\ref{Definitions.spin.zero.momentum}).
        From 
(\ref{Definitions.spin.zero.momentum})
        one can also see that contrary to the spin-independent case
(\ref{Definitions.scalar.zero.momentum}) there is no, in general, 
	factorization of the nuclear structure
        for $\sigma^A_{\rm SD}(0)$.
        Both proton $\langle{\bf S}^A_{p}\rangle$
        and neutron $\langle{\bf S}^A_{n}\rangle$
        spin contributions simultaneously enter into formula 
(\ref{Definitions.spin.zero.momentum})
        for the SD WIMP-nucleus cross section $\sigma^A_{\rm SD}(0)$.

        In the earlier considerations based on the OGM 
\cite{Engel:1989ix,Engel:1992bf} 
        one assumed that the nuclear spin is carried by the ``odd''
        unpaired group of protons or neutrons and only one of either 
        $\langle{\bf S}^A_n\rangle$ or $\langle{\bf S}^A_p\rangle$ 
        is non-zero (the same is true in the ISPSM 
\cite{Goodman:1985dc,Drukier:1986tm,Ellis:1988sh,Smith:1990kw}).
        In this case all possible target nuclei can naturally 
        be classified into n-odd and p-odd groups.

\begin{figure}[ht!] 
\begin{picture}(100,175)
\put(-40,-5){\includegraphics{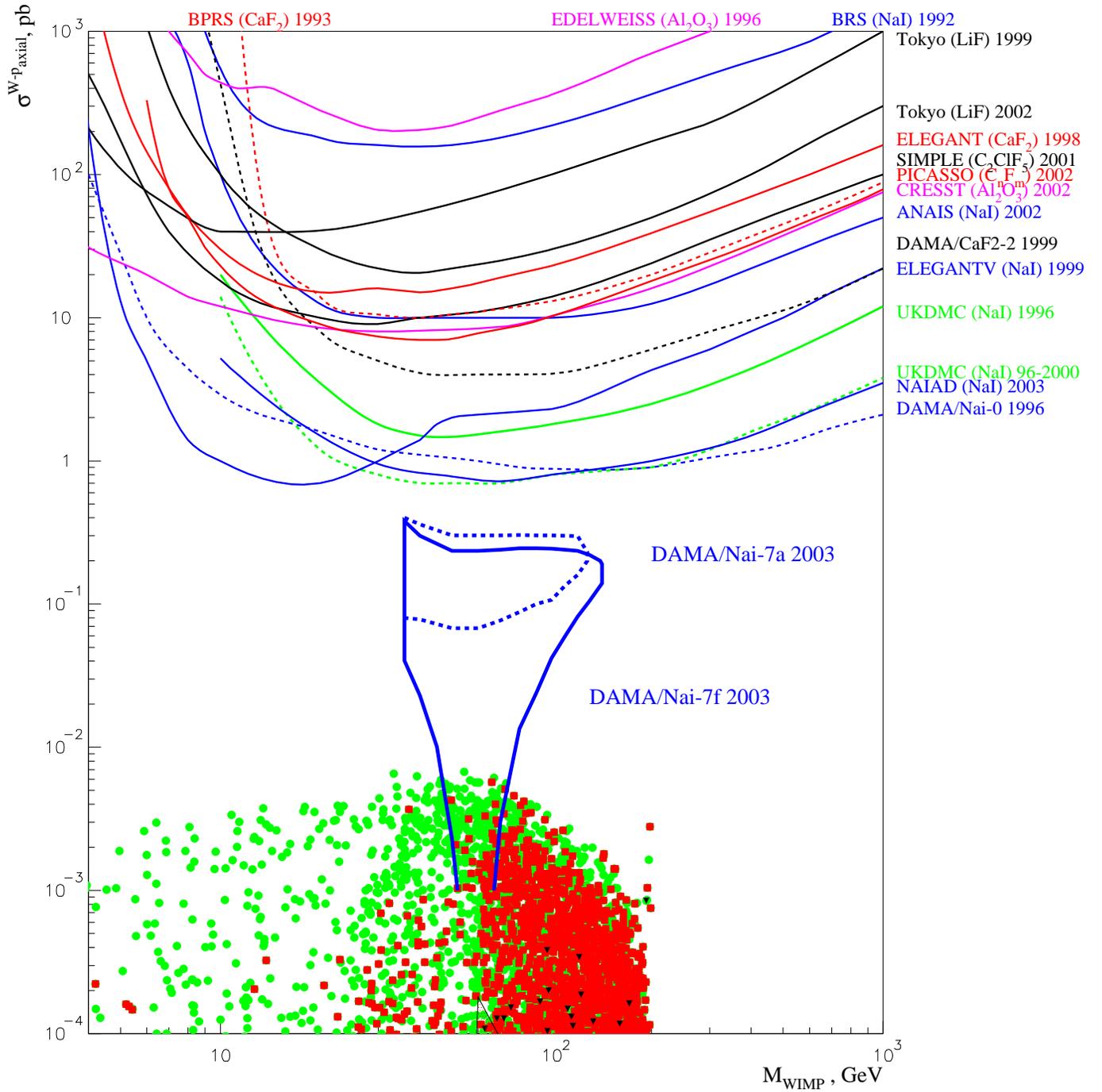}}
\end{picture}
\caption{Currently available exclusion curves for
        spin-dependent WIMP-proton cross sections
        ($\sigma^{p}_{{\rm SD}}$ 
	as a function of WIMP mass).
        The curves are obtained from   
\protect\cite{Bacci:1994ip,Bacci:1992pd,Bacci:1994tt,deBellefon:1996jh,%
Bernabei:1996vj,Bernabei:1997qb,Belli:1999xh,%
Sarsa:1996pa,Smith:1996fu,Sumner:1999yw,Spooner:2000kt,Ogawa:2000vi,%
Fushimi:1999kp,Yoshida:2000df,%
Ootani:1998jy,Minowa:1998ai,Ootani:1999pt,Ootani:1999xv,Miuchi:2002zp,%
Collar:2000zw,Angloher:2002in,Boukhira:2002qj,Cebrian:2002vd,Ahmed:2003su}.
	DAMA/NaI-7a(f) contours 
	for WIMP-proton SD interaction in $^{127}$I
	are obtained  
	on the basis of the positive and model independent 
	signature of annual signal modulation in the 
	framework of a mixed scalar-spin coupling approach
\cite{Bernabei:2003za,Bernabei:2001ve}. 
	The scatter plots correspond to our calculations given in 
Figs.~\ref{CrossSections-vs-lsp}--%
\ref{CrossSections-vs-mssm-spin}.
        The small triangle-like shaded area in the bottom 
	is taken from 
\cite{Ellis:2000ds}.
	Note that the {\it closed}\ DAMA contour is above the
	upper limit for $\sigma^p_{\rm SD} \approx 5\cdot 10^{-2}$ pb.
}
\label{Spin-p} 
\end{figure} 

\begin{figure}[!h] 
\begin{picture}(100,135)
\put(-30,-8){\includegraphics{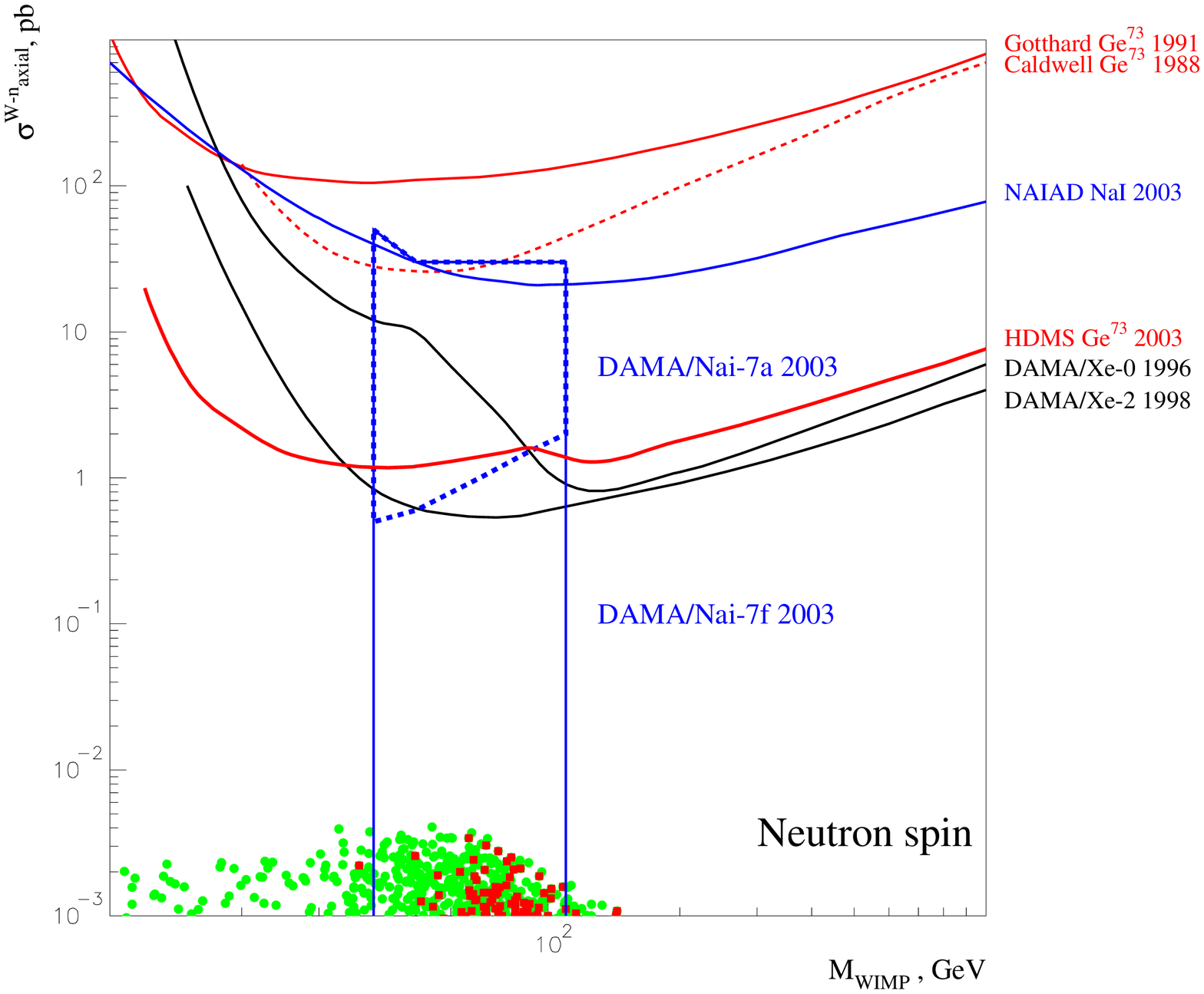}}
\end{picture}
\caption{Currently available exclusion curves for
        spin-dependent WIMP-neutron cross sections
        ($\sigma^{n}_{{\rm SD}}$ versus WIMP mass).
        The curves are taken from 
\protect\cite{Caldwell:1988su,Reusser:1991ri,%
Belli:1996sh,Belli:1996yf,Bernabei:1998ad,Ahmed:2003su,
Klapdor-Kleingrothaus:2002pi}.
	DAMA/NaI-7a(f) contours 
	for WIMP-neutron SD interaction
	(sub-dominating in $^{127}$I) 
	are obtained by us from the relevant figures of   
\cite{Bernabei:2003za,Bernabei:2001ve}. 
	The scatter plots correspond to our calculations given in 
Figs.~\ref{CrossSections-vs-lsp}--%
\ref{CrossSections-vs-mssm-spin}.
        Note that the NAIAD curve here corresponds to the 
	sub-dominant for $^{127}$I
	WIMP-neutron SD interaction. 
	The curve was extracted from the 
	nucleus $^{127}$I (which has dominating WIMP-proton SD interaction)
	in the approach of 
\cite{Tovey:2000mm}. It is much weaker in comparison 
	with the relevant NAIAD curve 
	for the WIMP-proton SD interaction in 
Fig.~\ref{Spin-p}.
}
\label{Spin-n} 
\end{figure} 
	
	Following this classification
        the current experimental situation
	in the form of the exclusion curves  
        for the spin-dependent WIMP-{\bf proton} cross section is given in 
Fig.~\ref{Spin-p}.      
        The data are taken from experiments BRS, (NaI, 1992)
\cite{Bacci:1994ip,Bacci:1992pd}, 
        BPRS (CaF$_2$, 1993)
\cite{Bacci:1994tt},
        EDELWEISS (sapphire, 1996)
\cite{deBellefon:1996jh}, 
        DAMA (NAI, 1996)
\cite{Bernabei:1996vj},
        DAMA (CaF$_2$, 1999)
\cite{Bernabei:1997qb,Belli:1999xh},
        UKDMS (NaI, 1996)
\cite{Sarsa:1996pa,Smith:1996fu,Sumner:1999yw,Spooner:2000kt},
        ELEGANT (CaF$_2$, 1998)
\cite{Ogawa:2000vi},
        ELEGANT (NaI, 1999)
\cite{Fushimi:1999kp,Yoshida:2000df},
        Tokio  (LiF, 1999, 2002)
\cite{Ootani:1998jy,Minowa:1998ai,Ootani:1999pt,Ootani:1999xv,Miuchi:2002zp},
        SIMPLE (C$_{2}$ClF$_{5}$, 2001)
\cite{Collar:2000zw},
        CRESST (Al$_2$O$_3$, 2002)
\cite{Angloher:2002in},
        PICASSO (C$_n$F$_m$, 2002)
\cite{Boukhira:2002qj}, 
        ANAIS (NaI, 2002) 
\cite{Cebrian:2002vd} and
        NAIAD (NaI, 2003)
\cite{Ahmed:2003su}.	
	Although the DAMA/NaI-7 (2003) contours 
\cite{Bernabei:2003za} are obtained  
	on the basis of the positive and model-independent 
	signature of the annual signal modulation (closed contour)
	as well as in the mixed coupling framework (open contour)
\cite{Bernabei:2001ve} 
	the contours for the WIMP-proton SD interaction 
	(dominating in $^{127}$I) 
	are also presented in the figure
	(we will discuss the situation later).

        The current experimental situation 
        for the spin-dependent WIMP-{\bf neutron} cross sections is given in 
Fig.~\ref{Spin-n}.
        The data are taken from the
        first experiments with natural Ge (1988, 1991)
\cite{Caldwell:1988su,Reusser:1991ri},
        xenon (DAMA/Xe-0,2)
\cite{Belli:1996sh,Belli:1996yf,Bernabei:1998ad}, 
        sodium iodide (NAIAD)
\cite{Ahmed:2003su}, 
	and from the HDMS experiment with a $^{73}${Ge} target  
\cite{Klapdor-Kleingrothaus:2002pi}.
	Similar to 
Fig.~\ref{Spin-p} the DAMA/NaI-7 (2003) 
\cite{Bernabei:2003za} 
	contours for the WIMP-neutron SD interaction
	(sub-dominant in $^{127}$I) are placed in the figure.
        In the future one can also expect some 
        exclusion curves for the SD cross section, for example, 
        from the CDMS 
\cite{Akerib:2003px}
	and EDELWEISS 
\cite{Benoit:2002hf} experiments
        with natural germanium bolometric detectors.

	In 
Figs.~\ref{Spin-p}~and~\ref{Spin-n}
	are also given scatter plots for SD proton and neutron
	cross sections 
	which correspond to the results of our calculations shown in 
Figs.~\ref{CrossSections-vs-lsp}--%
\ref{CrossSections-vs-mssm-spin}.
        From 
Figs.~\ref{Spin-p}~and~\ref{Spin-n}
	one can, in general, conclude that an  
        about two-orders-of-magnitude improvement 
        of the current DM experimental sensitivities 
	(in the form of these exclusion curves) is needed 
        to reach the SUSY predictions for the $\sigma^{p,n}_{{\rm SD}}$,  
        provided the SUSY lightest neutralino 
        is the best WIMP particle candidate. 

	Here we note that the calculated scatter plots for 
	$\sigma^p_{\rm SD}$ 
(Fig.~\ref{Spin-p})
	are obtained without any assumption about $\sigma^n_{\rm SD}=0$, 
        but the experimental exclusion curves for $\sigma^p_{\rm SD}$ 
	traditionally were extracted from the data 
	under the full neglection of the spin-neutron contribution,
	i.e. under the assumption $\sigma^n_{\rm SD}=0$.
	This one-spin-coupling dominance scheme (always used before
	a new approach was proposed in
\cite{Tovey:2000mm}) 
	gave a bit too pessimistic exclusion curves,
	but allowed direct comparison of exclusion curves 
	from different experiments.
        More stringent constraints on $\sigma^p_{\rm SD}$ 
	one obtains following 
\cite{Tovey:2000mm} and
\cite{Bernabei:2000qi,Bernabei:2003za,Bernabei:2003wy}
	assuming both 
	$\sigma^p_{\rm SD}\neq 0$ and $\sigma^n_{\rm SD}\neq 0$
	although usually the contribution of the neutron spin is
	very small
	($\langle{\bf S}^A_{n}\rangle \ll \langle{\bf S}^A_{p}\rangle$).
	Therefore the direct 
	comparison of old-fashioned exclusion curves 
	with new ones is misleading {\it in general}.
	The same conclusion concerns 
\cite{Bernabei:2003za,Bernabei:2003wy}
        direct comparison 
	of the SI exclusion curves 
	(obtained without any SD contribution) 
	with new SI exclusion curves 
	(obtained with non-zero SD contribution) 
	as well as with the results of the SUSY calculations
(Fig.~\ref{Scalar-2003}).

\subsection{Mixed spin-scalar WIMP-nucleon interactions}
        Further more accurate calculations of spin nuclear structure 
\cite{Iachello:1991ut,Engel:1991wq,Pacheco:1989jz,%
      Engel:1992qb,Engel:1995gw,Dimitrov:1995gc,%
      Ressell:1993qm,Ressell:1997kx,%
      Vergados:1996hs,Kosmas:1997jm,Divari:2000dc}
        demonstrate that contrary to the simplified odd-group approach both
        $\langle{\bf S}^A_{p}\rangle$ and $\langle{\bf S}^A_{n}\rangle$ 
        differ from zero, but nevertheless 
        one of these spin quantities always dominates
        ($\langle{\bf S}^A_{p}\rangle \ll \langle{\bf S}^A_{n}\rangle$, or
         $\langle{\bf S}^A_{n}\rangle \ll \langle{\bf S}^A_{p}\rangle$). 
{\it If together}\ with the dominance like 
        $\langle{\bf S}^A_{p(n)}\rangle \ll \langle{\bf S}^A_{n(p)}\rangle$
        one would have the WIMP-proton and WIMP-neutron couplings
        of the same order of magnitude
        ({\it not} $a_{n(p)}\! \ll\! a_{p(n)}$), 
        the situation could look like that in the odd-group model
	and one could safely (at the current level of accuracy)
	neglect sub-dominant spin contribution in the data analysis.

        Nevertheless it was shown in 
\cite{Tovey:2000mm}
        that in the general SUSY model one can meet right 
	a case when 
        $a_{n(p)}\! \ll\! a_{p(n)}$
	and proton and neutron spin contributions are strongly mixed.    
        To separately constrain the SD proton and neutron contributions 
	at least two new approaches appeared
	in the literature
\cite{Tovey:2000mm,Bernabei:2001ve}.
        As the authors of 
\cite{Tovey:2000mm} claimed, their method 
        has the advantage that the limits on individual 
        WIMP-proton and WIMP-neutron SD cross sections 
        for a given WIMP mass can be combined 
        to give a model-independent limit 
        on the properties of WIMP scattering 
        from both protons and neutrons in the target nucleus. 
        The method relies on the assumption that the 
        WIMP-nuclear SD cross section can be presented in the form
$\displaystyle 
        \sigma^A_{\rm SD}(0) \!=\!
        \left( \sqrt{\sigma^p_{\rm SD}|^{}_A} \pm 
       \sqrt{\sigma^n_{\rm SD}|^{}_A} \right)^2$,
        where $\sigma^p_{\rm SD}|^{}_A$
          and $\sigma^n_{\rm SD}|^{}_A$ are auxiliary 
        quantities, not directly connected with measurements.
        Furthermore, to extract, for example, a constraint on
        the sub-dominant WIMP-proton spin 
        contribution one should assume the proton contribution dominance 
        for a nucleus whose spin is almost completely 
        determined by the neutron-odd group. 
	From one side,
	this may look almost useless, 
        especially because these sub-dominant 
        constraints are always much weaker 
        than the relevant constraints 
        obtained directly with a proton-odd group target
	(one can compare, for example, the
	restrictive potential of the NAIAD exclusion curves in 
Figs.~\ref{Spin-p} and \ref{Spin-n}). 
	From another side, the 
        very large and very small ratios $\sigma_p/\sigma_n
	\sim a_{p}/a_{n}$ obtained in 
\cite{Tovey:2000mm} correspond to neutralinos which are
	extremely pure gauginos. 
	In this case $Z$-boson exchange in SD interactions is absent 
	and only sfermions give contributions to SD cross sections. 
	Obviously this is a very particular case
	which is also currently not in agreement with 
	the experiments.
	Following an analogy between neutrinos and neutralinos
	on could assume that neutralino couplings
	with the neutron and the proton should not to very different
\cite{Vergados:2003dubna}
	and one could expect preferably 
        $|a_n|/|a_p| \approx O(1)$.
	We have checked the assumption in our effMSSM approach 
	for large LSP masses in
\cite{Bednyakov:2002dz,Bednyakov:2003wf} 
	and in this
	paper for relatively low LSP masses 
	$m_\chi < 200$~GeV.
\begin{figure}[!h] 
\begin{picture}(100,110)
\put(-18,-6){\includegraphics{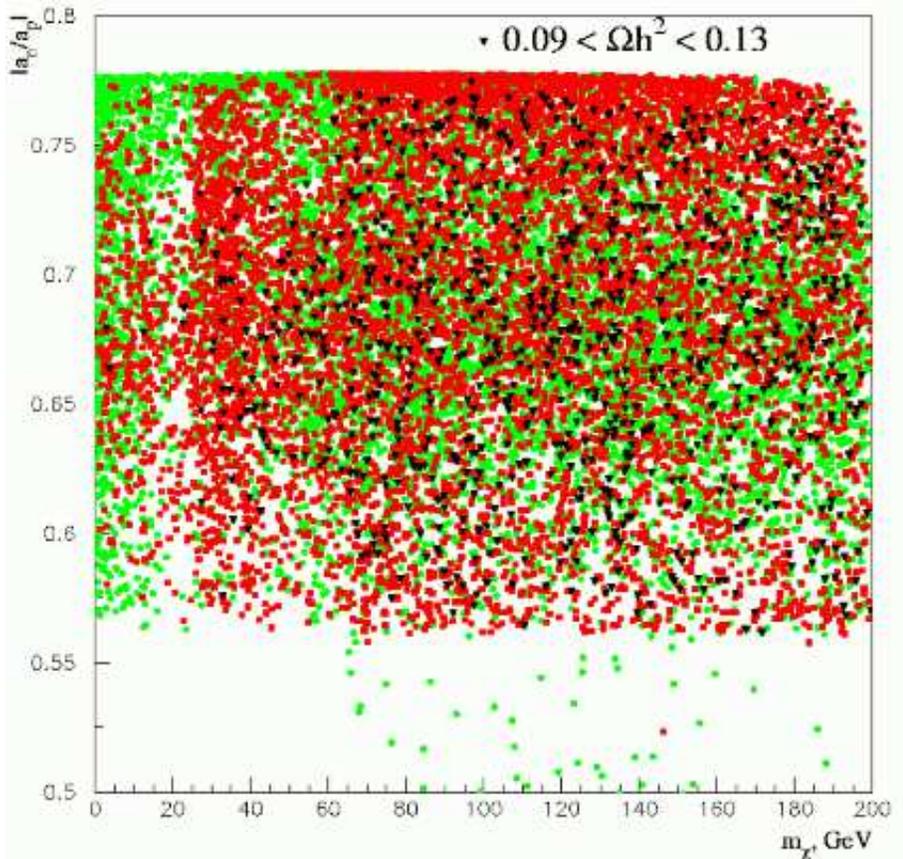}}
\end{picture} 
\caption{The scatter plots (circles, squares and triangles) 
	give the ratio of the neutralino-neutron spin coupling
        $a_n$ to the neutralino-proton spin coupling $a_p$ 
	in the effMSSM 
	under the notations as in 
Figs.~\ref{CrossSections-vs-lsp}--%
\ref{CrossSections-vs-mssm-spin}. 
	The ratio is restricted to the range between 
	0.55 and 0.8. 
}
\label{an2ap-ratio}
\end{figure} 
Figure~\ref{an2ap-ratio} shows that 
	for the ratio of $a_n$ to $a_p$ we have the bounds
\begin{equation}
\label{an2ap-ratio-bounds}
0.55 < \left|\frac{a_n}{a_p} \right| <  0.8.
\end{equation}
        The scatter plots in 
Fig.~\ref{an2ap-ratio} as previously 
(see Fig.~\ref{CrossSections-vs-lsp}) were obtained with the 
	relic neutralino density  
	$0.0< \Omega_\chi h^2_0<1.0$ (green circles),
	with sub-dominant relic neutralino contribution 
 	$0.002 < \Omega_\chi h^2_0<0.1$ (red squares) 
	and with a WMAP-inspired relic neutralino density of
	$0.094 < \Omega_\chi h^2_0<0.129$ (black triangles).
        Therefore in the model the couplings are almost the same
        and one can safely neglect, for example, 
	the $\langle{\bf S}^A_{p}\rangle$-spin 
        contribution in the analysis of the 
	DM data for a nuclear target with 
	$\langle{\bf S}^A_{p}\rangle \ll 
         \langle{\bf S}^A_{n}\rangle$.

	Furthermore, when one compares in the same figure
	an exclusion curve for SD WIMP-proton coupling
	obtained without sub-dominant SD WIMP-neutron contribution 
	and without SI contribution (all curves in 
Fig.~\ref{Spin-p} except the one for NAIAD
\cite{Ahmed:2003su} and one for Tokyo-LiF
\cite{Miuchi:2002zp}), 
	with a curve from the approach of 
\cite{Tovey:2000mm}, when the sub-dominant contribution is included
	(the NAIAD and Tokyo-LiF curves in 
Fig.~\ref{Spin-p}),
	one {\it ``artificially''}\ improves the sensitivity 
	of the {\it latter}\ curves 
	(NAIAD or Tokyo-LiF) in comparison with the former ones.
	To be consistent and for reliable comparisons, 
	one should coherently recalculate
	all previous curves in the new manner. 
	This message is clearly also stressed in 
\cite{Bernabei:2003za}. 
	The same arguments are true for the last results 
	of the SIMPLE experiment 
\cite{Giuliani:2003nf} and search for DM with NaF bolometers   
\cite{Takeda:2003km} where the SI contribution seems also completely ignored.
	Both above-mentioned results for fluorine 
	will obviously be worse if (contrary to calculations of 
\cite{Pacheco:1989jz}) more reliable $^{19}$F spin matrix elements 
	(for example, from 
\cite{Divari:2000dc}) 
	were used in their analysis. 
	Although $^{19}$F has the best properties 
	for investigation of WIMP-nucleon spin-dependent interactions  
(see, for example
\cite{Divari:2000dc})
	it is not obvious that one should completely ignore
	spin-independent WIMP coupling with the fluorine. 
	For example, in the relation  
$\sigma^A \sim \sigma^{A,p}_{\rm SD}
         \left[\frac{\sigma^A_{\rm SI}}{\sigma^{A,p}_{\rm SD}}
        +\left(1 + 
	\sqrt{\frac{\sigma^{A,n}_{\rm SD}}{\sigma^{A,p}_{\rm SD}}}
        \right)^2 
        \right]
$	which follows from 
(\ref{Definitions.scalar.zero.momentum})--%
(\ref{Definitions.spin.zero.momentum.Bernabei}),
	it is not a priori clear that 
        $\frac{\sigma^A_{\rm SI}}{\sigma^{A,p}_{\rm SD}} \ll
	\frac{\sigma^{A,n}_{\rm SD}}{\sigma^{A,p}_{\rm SD}}
	$.
	At least for isotopes with atomic number $A>50$
\cite{Bednyakov:1994qa,Jungman:1996df}
	the neglection of the SI contribution would be a larger 
        mistake than the neglection of the  
        sub-dominant SD WIMP-neutron contribution,  
	when the SD WIMP-proton interaction dominates.
	Therefore we would like to note that 
	the ``old'' odd-group-based approach in 
	analyzing the SD data from experiments with heavy enough
	targets (for example, germanium) is still quite suitable. 
	Especially when it is not obvious that 
	(both) spin couplings dominate over the scalar one.

        From measurements with $^{73}$Ge one can extract, in principle, 
        not only the dominant constraint for WIMP-nucleon coupling
        $a_n$ (or $\sigma_{\rm SD}^{n}$) 
        but also the constraint for the sub-dominant WIMP-proton coupling
        $a_p$ (or $\sigma_{\rm SD}^{p}$) using the approach of 
\cite{Tovey:2000mm}.
        Nevertheless, the latter constraint will be much weaker
        in comparison with the constraints from p-odd group
        nuclear targets, like $^{19}$F or I.  
        This fact is illustrated by the NAIAD (NaI, 2003) curve in 
Fig.~\ref{Spin-n}, which corresponds to the sub-dominant
        WIMP-neutron spin contribution 
        extracted from the p-odd nucleus I.

\smallskip
         Another approach of Bernabei et al.
\cite{Bernabei:2001ve} looks in a more 
	appropriate way 
	for the mixed spin-scalar coupling data presentation,
	and is based on an introduction of the so-called effective 
	SD nucleon cross section $\sigma^{pn}_{\rm SD}(0)$
	(originally $\sigma_{{\rm SD}}$ in 
\cite{Bernabei:2003za,Bernabei:2001ve}) 
	and coupling mixing angle $\theta$
(\ref{effectiveSD-cs})
	instead of $\sigma^{p}_{\rm SD}(0)$ and
	$\sigma^{n}_{\rm SD}(0)$.
        With these definitions the SD WIMP-proton and
        WIMP-neutron cross sections are given by relations
(\ref{effectiveSD-cs-pn}).

\begin{figure}[!h] 
\begin{picture}(100,80)
\put(-35,0){\includegraphics{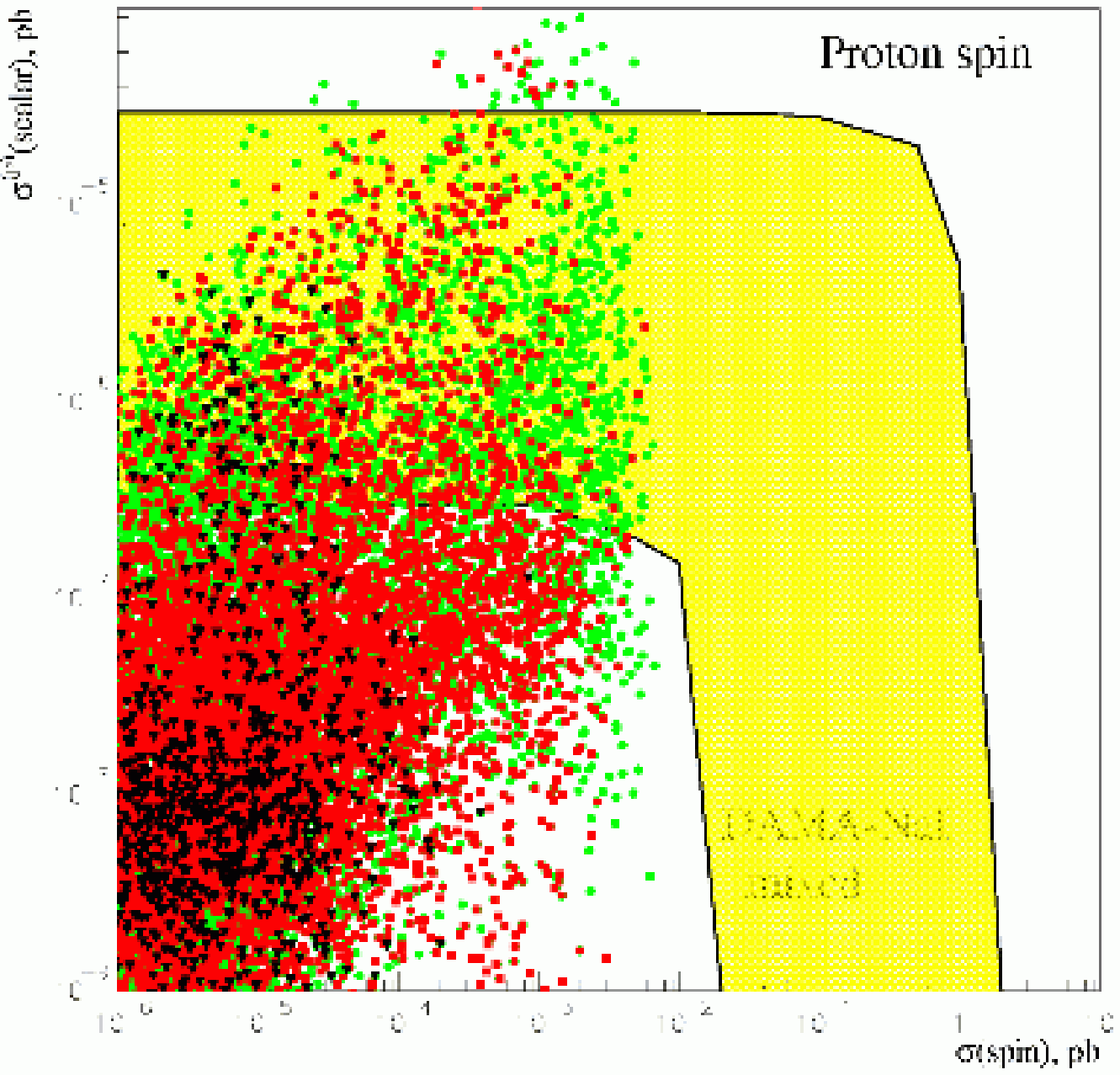}}
\put( 53,0){\includegraphics{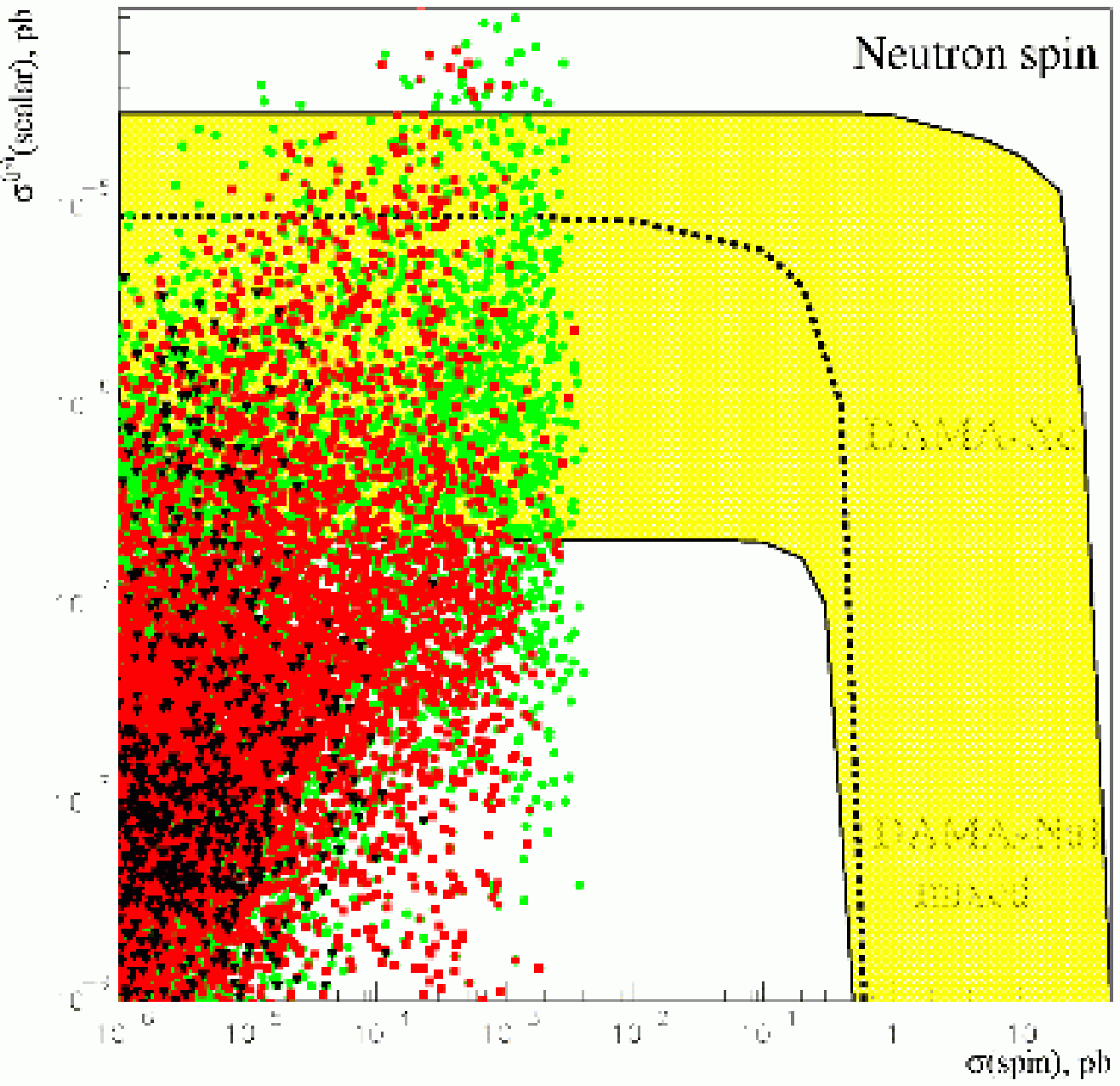}}
\end{picture}
\caption{ 
        The DAMA-NaI region 
        from the WIMP annual modulation signature 
        in the ($\xi \sigma_{\rm SI}$, $\xi \sigma_{\rm SD}$) space 
        for $40<m^{}_{\rm WIMP}<110$~GeV
\cite{Bernabei:2003za,Bernabei:2001ve}.
	Left panel corresponds to
	dominating (in $^{127}$I)  
	SD proton coupling only ($\theta$ = 0)  
	and right panel corresponds to 
	sub-dominating SD neutron coupling only 
        ($\theta$ = $\pi/2$). 
        The scatter plots give correlations
        between $\sigma^{p}_{{\rm SI}}$ and 
        $\sigma^{}_{{\rm SD}}$ in the effMSSM ($\xi=1$ is assumed) 
	for $m_\chi<200$~GeV under the same notations as in 
Figs.~\ref{CrossSections-vs-lsp}--%
\ref{CrossSections-vs-mssm-spin}. 
	In the right panel the DAMA liquid xenon exclusion curve from 
\cite{Bernabei:2001ve} is given (dashed line).}
\label{Bernabei:2001ve:fig}
\end{figure} 
In Fig.~\ref{Bernabei:2001ve:fig}
        the WIMP-nucleon spin and scalar mixed couplings 
        allowed by the annual modulation signature from 
        the 100-kg DAMA/NaI experiment 
        are shown inside the shaded regions. 
        The regions from 
\cite{Bernabei:2003za,Bernabei:2001ve} 
        in the ($\xi \sigma_{\rm SI}$, $\xi \sigma_{\rm SD}$) space 
        for 40~GeV$<m^{}_{\rm WIMP}<$110~GeV cover
	spin-scalar mixing coupling for the proton ($\theta=0$ case of 
\cite{Bernabei:2003za,Bernabei:2001ve}, left panel) and 
	spin-scalar mixing coupling for the neutron 
	($\theta$ = $\pi/2$, right panel). 
	From nuclear physics one has for the proton spin dominated
        $^{23}$Na and $^{127}$I \
        $\frac{\langle {\bf S}_n\rangle}{\langle {\bf S}_p\rangle}< 0.1$ 
	and 
        $\frac{\langle {\bf S}_n\rangle}
	{\langle {\bf S}_p\rangle}< 0.02 \div 0.23$,
	respectively.
        For the $\theta=0$ due to the p-oddness of the I target, 
	the DAMA WIMP-proton spin constraint is the most severe one
(see Fig.~\ref{Spin-p}).  

	In the right panel of  
Fig.~\ref{Bernabei:2001ve:fig}
	we present the exclusion curve (dashed line) for the
	WIMP-proton spin coupling from the proton-odd isotope 
	$^{129}$Xe obtained under the mixed coupling assumptions  
\cite{Bernabei:2001ve} from the DAMA-LiXe (1998) experiment 
\cite{Bernabei:2002qg,Bernabei:1998ad,Bernabei:2002af}.
	For the DAMA NaI detector the 
	$\theta=\pi/2$ means no 
${\langle {\bf S}_p\rangle}$ contribution at all. 
	Therefore, in this case DAMA gives the 
        sub-dominant ${\langle {\bf S}_n\rangle}$ 
	contribution only,
        which could be compared further with the dominant 
        ${\langle {\bf S}_n\rangle}$ contribution in $^{73}$Ge.
	
  	The scatter plots in
Fig.~\ref{Bernabei:2001ve:fig} give 
	$\sigma^{p}_{{\rm SI}}$ as a function of 
	$\sigma^p_{{\rm SD}}$ (left panel) and  
	$\sigma^n_{{\rm SD}}$  (right panel)
	calculated in this work in the effMSSM with parameters from 
(\ref{Scanning})
	under the same constraints on the 
	relic neutralino density as in 
Figs.~\ref{CrossSections-vs-lsp}--%
\ref{CrossSections-vs-mssm-spin}.
	Filled circles (green) correspond to relic neutralino density 
	$0.0< \Omega_\chi h^2_0<1.0$,
	squares (red) correspond to sub-dominant relic neutralino
	contribution 
 	$0.002 < \Omega_\chi h^2_0<0.1$ 	
	and triangles (black)
	correspond to 
	WMAP density constraint 
	$0.094 < \Omega_\chi h^2_0<0.129$. 

	The constraints on the SUSY parameter space in the mixed coupling 
	framework in Fig.~\ref{Bernabei:2001ve:fig} 
	are, in general, 
	much stronger in comparison with the 
	traditional approach based on the one-coupling dominance
(Figs.~\ref{Scalar-2003}, \ref{Spin-p} and \ref{Spin-n}).

	It follows from 
Fig.~\ref{Bernabei:2001ve:fig}, 
	that when the LSP is the sub-dominant DM particle (squares in the
	figure) SD WIMP-proton and WIMP-neutron cross sections 
	at a level of $3\div5\cdot 10^{-3}$~pb are allowed,
	but the WMAP relic density constraint (triangles)
	together with the DAMA restrictions leaves only 
	$\sigma_{\rm SD}^{p,n}<3\cdot 10^{-5}$~pb
	without any visible reduction of allowed values for
	$\sigma^p_{\rm SI}$.
	In general, according to the DAMA restrictions, 
	small SI cross sections are completely excluded, 
	only $\sigma^p_{\rm SI}> 3\div5\cdot 10^{-7}$~pb are allowed. 
	Concerning the SD cross section the situation is not clear,
	because for the allowed values of the SI contribution,  
	the SD DAMA sensitivity did not yet reach  
	the calculated upper bound for the SD LSP-proton
	cross section of $5\cdot 10^{-2}$~pb.

\subsection{The mixed couplings case for the high-spin \boldmath $^{73}$Ge}
        Comparing the number of exclusion curves in 
Figs.~\ref{Spin-p} and \ref{Spin-n} 
        one can see that there are many measurements with p-odd nuclei
        and there is a lack of data for n-odd nuclei, i.e. 
        for $\sigma_{\rm SD}^{n}$. 
        Therefore measurements with  n-odd nuclei are needed.
        From our point of view this lack of $\sigma_{\rm SD}^{n}$ 
        measurements can be 
        filled with new data expected from the HDMS experiment with 
        the high-spin isotope $^{73}$Ge
\cite{Klapdor-Kleingrothaus:2002pg}.
        This isotope looks with a good accuracy 
        like an almost pure n-odd group nucleus with 
        $\langle {\bf S}_{n}\rangle\! \gg\! \langle {\bf S}_{p}\rangle$
(Table~\ref{Nuclear.spin.main.table.71-95}).
        The variation of the $\langle {\bf S}_{p}\rangle$
        and $\langle {\bf S}_{n}\rangle$
        in the table reflects the level of  
        inaccuracy and complexity 
        of the current nuclear structure calculations.
\begin{table}[h!] 
\caption{All available calculations in different nuclear models for the
	zero-momentum spin structure (and predicted magnetic moments $\mu$) 
        of the $^{73}$Ge nucleus. 
        The experimental value of the magnetic moment given in the brackets 
        is used as input in the calculations.
\label{Nuclear.spin.main.table.71-95}}
\begin{center}
\smallskip
\begin{tabular}{lrrr}
\hline\hline
$^{73}$Ge~($G_{9/2}$) & ~~~~~~~~$\langle {\bf S}_p \rangle$ & 
~~~~~~~~$\langle {\bf S}_n \rangle$ & ~~~~~~~~$\mu$ (in $\mu_N$) \\ \hline
ISPSM, Ellis--Flores~\cite{Ellis:1988sh,Ellis:1991ef}
        &    0    & $0.5$               & $-1.913$ \\ 
OGM, Engel--Vogel~\cite{Engel:1989ix}   
        &    0    & $0.23$      &$(-0.879)_{\rm exp}$ \\ 
IBFM, Iachello at al.~\cite{Iachello:1991ut} and \cite{Ressell:1993qm}
        &$-0.009$ & $0.469$ &$-1.785$\\ 
IBFM (quenched), 
        Iachello at al.~\cite{Iachello:1991ut} and \cite{Ressell:1993qm}
        &$-0.005$  & $0.245$ &$(-0.879)_{\rm exp}$ \\
TFFS, Nikolaev--Klapdor-Kleingrothaus, \cite{Nikolaev:1993dd} 
        &$0$   & $0.34$ & --- \\ 
SM (small), Ressell at al.~\cite{Ressell:1993qm} 
        &$0.005$   & $0.496$ &$-1.468$ \\ 
SM (large), Ressell at al.~\cite{Ressell:1993qm} 
        &$0.011$   & $0.468$ &$-1.239$ \\ 
SM (large, quenched), Ressell at al.~\cite{Ressell:1993qm} 
        &$0.009$   & $0.372$ &$(-0.879)_{\rm exp}$ \\ 
``Hybrid'' SM, Dimitrov at al.~\cite{Dimitrov:1995gc}           
        & $0.030$ & $0.378$ & $-0.920$ \\ 
\hline\hline
\end{tabular} \end{center}
\end{table} 

	In the mixed spin-scalar coupling case for $^{73}$Ge
	the direct detection rate integrated over recoil energy 
(\ref{for-toy-mixig}) from threshold energy, $\eth$, 
	till maximal energy, $\emx$ 
	can be presented in the form 
\begin{eqnarray}
\label{for-Ge-73}
R(\eth, \emx)&=&
	\alpha(\eth,\emx,m_\chi)\,\sigma^p_\SI
	+\beta(\eth,\emx,m_\chi)\,\sigma^{n}_\SD;\\
&&\alpha(\eth,\emx,m_\chi)
	= N_T \frac{\rho_\chi M_A}
		{2 m_\chi \mu_p^2 } A^2 
	A_\SI(\eth,\emx),\nonumber \\
&& \nonumber
\beta(\eth,\emx,m_\chi)
 	=	N_T \frac{\rho_\chi M_A} {2 m_\chi \mu_p^2 } 
	\frac43 \frac{J+1}{J}
	\langle {\bf S}^A_n\rangle^2 A_\SD(\eth,\emx). 
\end{eqnarray}
	The convolutions of nuclear form-factors with the WIMP 
	velocity distributions, 	
	$A_{\SI,\SD}(\eth,\emx)$, are defined by expressions 
(\ref{for-toy-mixig-FF}).
	We neglect  for $^{73}${Ge} the sub-dominant contribution from 
	WIMP-proton spin coupling proportional 
	to $\langle {\bf S}^A_p\rangle$.
	We consider only a simple spherically symmetric isothermal 
	WIMP velocity distribution 
\cite{Drukier:1986tm,Freese:1988wu}
	and do not go into details of any possible
	and in principle important 
	uncertainties (and/or modulation effects) 
	of the Galactic halo WIMP distribution 
\cite{Kinkhabwala:1998zj,Donato:1998pc,Evans:2000gr,%
Green:2000jg,Copi:2000tv,Ullio:2000bf,Vergados:2000cp}.
	For simplicity we use the gaussian 
	scalar and spin nuclear form-factors from 
\cite{Ellis:1993ka,Ellis:1991ef}.
	For the relic neutralino 
	mass density and for the escape neutralino velocity 
	we use the values 0.3~GeV$/$cm$^3$ and 600 km$/$s, 
	respectively. 
	With formulas 
(\ref{for-Ge-73}), we perform below a simple estimation of prospects
	for DM search and SUSY constraints with the high-spin $^{73}$Ge
	detector HDMS assuming mixing of WIMP-neutron spin and 
	WIMP-nucleon scalar couplings together with 
	available results from the 
	DAMA-NaI and LiXe experiments
\cite{Bernabei:2000qi,Bernabei:2003za,Bernabei:2003wy,%
      Bernabei:2002qg,Bernabei:1998ad,Bernabei:2002af}.

	The Heidelberg Dark Matter Search (HDMS) experiment
	uses a special configuration of 
	two Ge detectors to efficiently reduce the background
\cite{Klapdor-Kleingrothaus:2002pg,Klapdor-Kleingrothaus:2000uh}.
	From the first preliminary
        results of the HDMS experiment 
	with inner HPGe crystal of enriched $^{73}$Ge
\cite{Klapdor-Kleingrothaus:2002pg,Klapdor-Kleingrothaus:2002pi}
	we can estimate the current background event rate 
	$R(\eth, \emx)$ integrated here from  
	the ``threshold'' energy $\eth=15$~keV to ``maximal'' 
	energy $\emx=50$~keV. 
	We obtain $R(15,50)\approx 10$ events/kg/day.
	A substantial improvement of the background 
	(up to an order of magnitude) is further expected for
	the setup in the Gran Sasso Underground Laboratory.
\begin{figure}[!h] 
\begin{picture}(100,100)
\put(-10,-5){\includegraphics{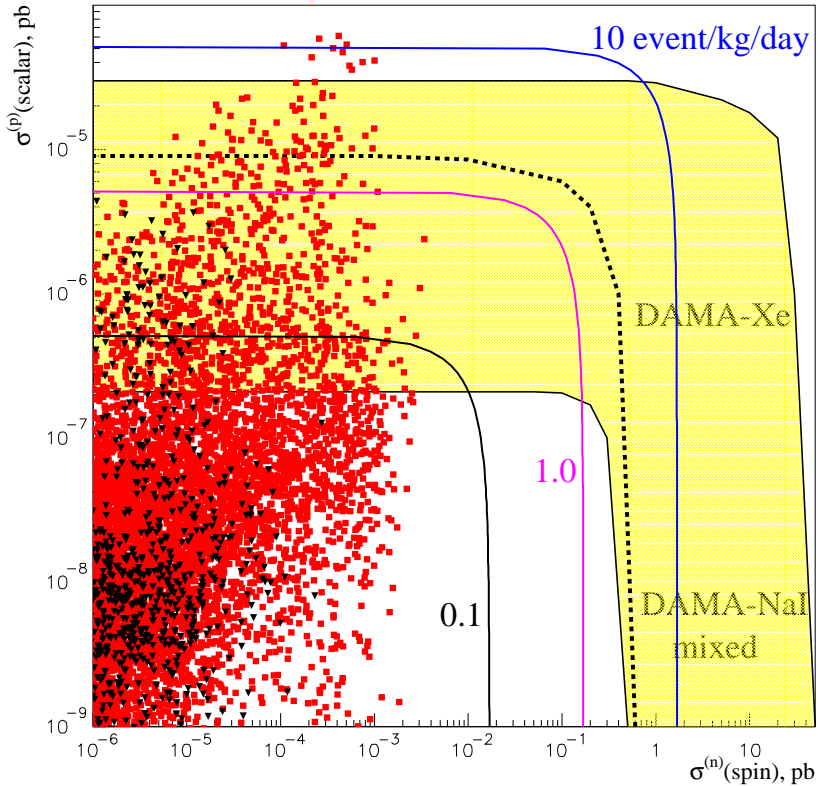}}
\end{picture}
\caption{ 
	The solid lines (marked with numbers of $R(15,50)$ in events/kg/day)
	show the sensitivities of the HDMS setup with $^{73}$Ge
	in the framework of mixed SD WIMP-neutron 
	and SI WIMP-nucleon couplings.
        The DAMA-NaI region for sub-dominant SD WIMP-neutron coupling 
        ($\theta$ = $\pi/2$) is from 
Fig.~\ref{Bernabei:2001ve:fig}.
        Scatter plots give correlations
        between $\sigma^{p}_{{\rm SI}}$ and 
        $\sigma^{n}_{{\rm SD}}$ in the effMSSM 
	for $m_\chi<200$~GeV.
	The squares (red) correspond to sub-dominant relic neutralino
	contribution $0.002 < \Omega_\chi h^2_0<0.1$ 	
	and triangles (black)
	correspond to WMAP relic neutralino density 
	$0.094 < \Omega_\chi h^2_0<0.129$. 
	The dashed line from 
\cite{Bernabei:2001ve}
	shows the DAMA-LiXe (1998) exclusion curve 
	for $m^{}_{\rm WIMP}=50$~GeV. }
\label{hdms-toy}
\end{figure} 
In Fig.~\ref{hdms-toy}
	solid lines for the integrated rate $R(15,50)$
	marked with numbers 10, 1.0 and 0.1 (in events/kg/day) 
	present for $m^{}_{\rm WIMP}=70$~GeV
	our exclusion curves expected from the HDMS setup with $^{73}$Ge
	in the framework of mixed SD WIMP-neutron 
	and SI WIMP-nucleon couplings.
	Unfortunately the current background index 
	for HDMS is not yet optimized,
	and the relevant 
	exclusion curve (marked with 10 events/kg/day) 
	has almost the same strength to reduce $\sigma^{n}_{{\rm SD}}$ 
	as the dashed curve from the DAMA experiment with liquid Xe
\cite{Bernabei:2001ve} obtained for $m^{}_{\rm WIMP}=50$~GeV
	(better sensitivity is expected with HDMS for 
	$m^{}_{\rm WIMP}<40$~GeV). 
	However, both experiments lead already to some sharper restriction 
	for $\sigma^{n}_{\SD}$ then obtained by DAMA (see 
Fig.~\ref{hdms-toy}). 
	One order of magnitude improvement of the 
	HDMS sensitivity (curve marked with 1.0)
	will supply us with the best exclusion curve for 
	SD WIMP-neutron coupling, but this sensitivity is not
	yet enough to reach the calculated 
	upper bound for $\sigma^{n}_{{\rm SD}}$.   
	This sensitivity also could reduce the 
	upper bound for SI WIMP-proton coupling $\sigma^{p}_{{\rm SI}}$
	to a level of $10^{-5}$~pb.
	Nevertheless, only an {\it additional} about-one-order-of-magnitude
	HDMS sensitivity improvement is needed to obtain
	decisive constraints on 
	$\sigma^{p}_{{\rm SI}}$ as well as on
	$\sigma^{n}_{{\rm SD}}$.
	In this case only quite narrow bounds  
	for these cross sections will be allowed
	(below the curve marked by 0.1 and above the 
	lower bound of DAMA-NaI mixed region). 
	In practice it seems, that only the 
	DAMA and the HDMS constraints {\it together}\ could 
	restrict the SD WIMP-neutron coupling sufficiently.

\subsection{Some other consequences of the DAMA results}
	It follows from 
Figs.~\ref{Scalar-2003},
\ref{Spin-p},
\ref{Spin-n} and
\ref{Bernabei:2001ve:fig}  
	that the main results of the DAMA experiment 
	one could summarize in the limitations of the WIMP mass, 
	and the restrictions on the 
	cross section of the 
{\em scalar}\ WIMP-proton interaction.
\begin{figure}[!h]  
\begin{picture}(100,75)
\put(-30,-78){\includegraphics{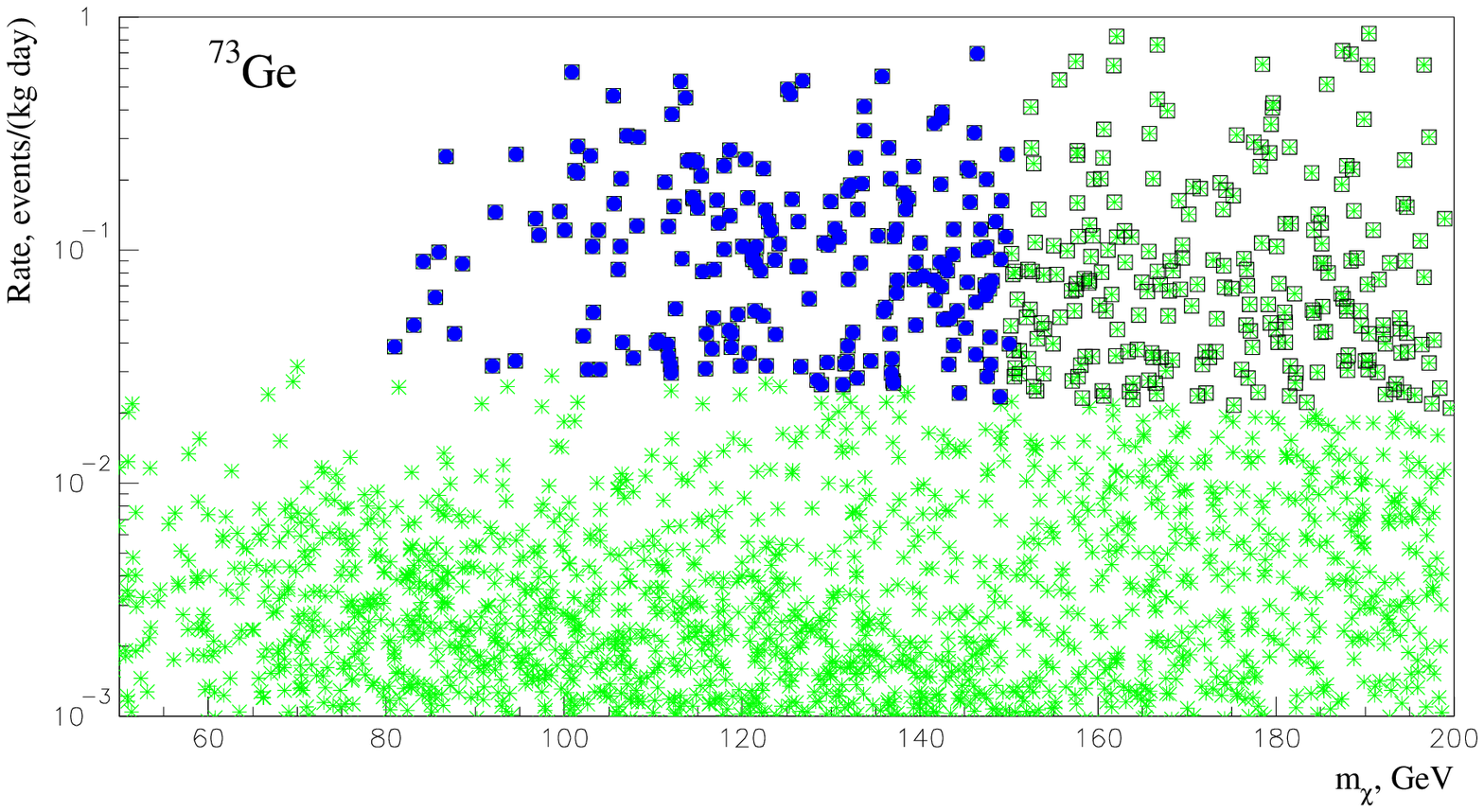}}
\end{picture}
\caption{Event rate for direct neutralino detection in a 
	$^{73}$Ge detector as function of the LSP neutralino mass. 
	Green crosses present our calculations with relic density constraint
	$0.1 < \Omega_\chi h^2_0<0.3$ only.
	Open boxes correspond to implementation of the 
	SI cross section limit
	$1\cdot10^{-7}~{\rm pb}<\sigma^p_\SI(0)<3\cdot 10^{-5}~{\rm pb}$ 
	only, and closed boxes show results
	with the additional WIMP-mass constraint 
	$40 < m^{}_{\rm WIMP} < 150~{\rm GeV}$
(see (\ref{DAMA-main-we})). 
\label{RateGe-damarg}}
\begin{picture}(100,70)
\put(-30,-84){\includegraphics{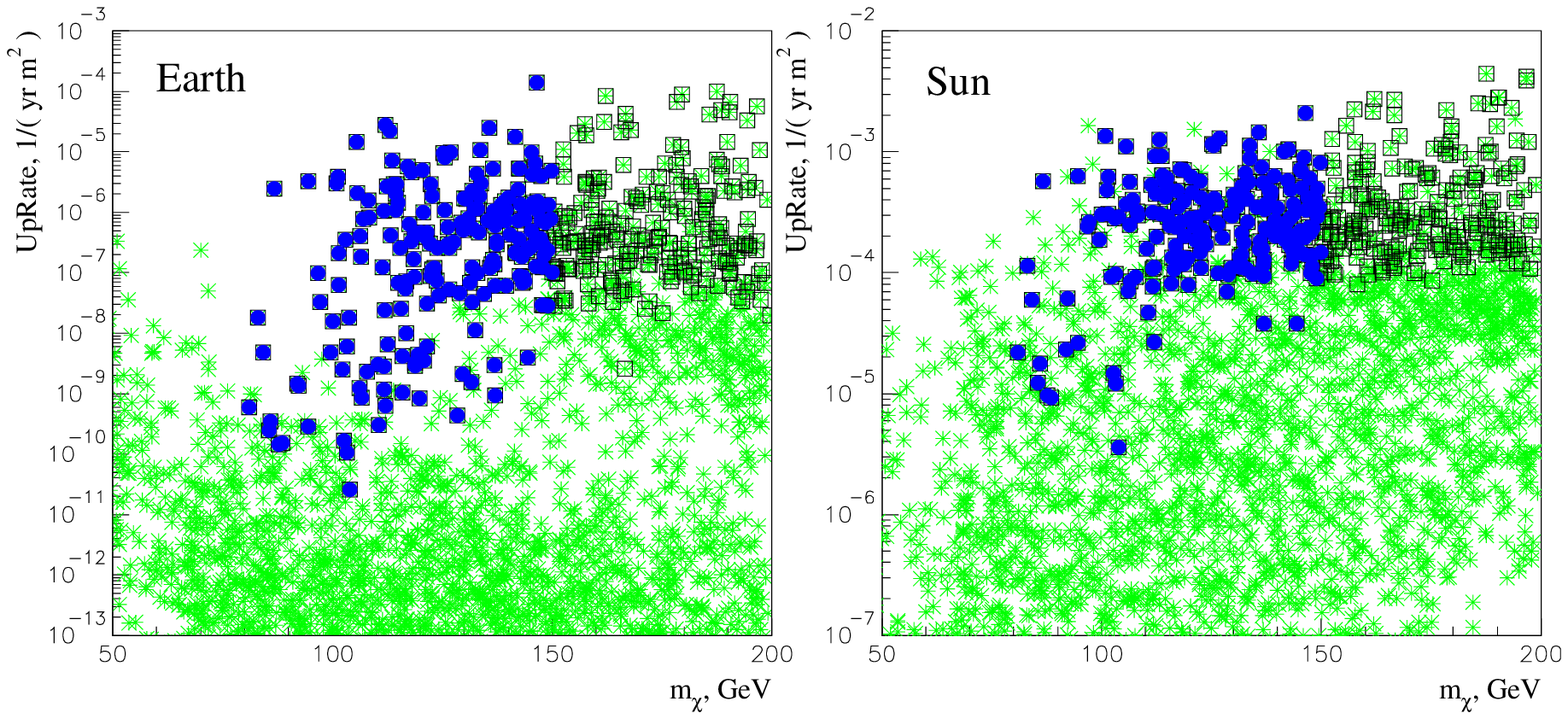}}
\end{picture}
\caption{Indirect detection rate for upgoing muons 
	from DM particles (neutralinos) 
	annihilation in the Earth (a) and the Sun (b)
	as function of the LSP neutralino mass. 
	Green crosses present our calculations with relic density constraint
	$0.1 < \Omega_\chi h^2_0<0.3$ only.
	Open boxes correspond to implementation of the 
	SI cross section limits of
	(\ref{DAMA-main-we}) only and closed boxes depict results
	with both limitations of 
(\ref{DAMA-main-we}). 
\label{Upmuons-damarg}
}
\end{figure} 
	Quite approximately (having in mind all possible uncertainties of 
\cite{Bernabei:2003za,Bernabei:2003wy}) 
	one can write them in the form:
\begin{equation}
\label{DAMA-main-we}
40~{\rm GeV} < m^{}_{\rm WIMP} < 150~{\rm GeV}, \quad 
1\cdot10^{-7}~{\rm pb}<\sigma^p_\SI(0)<3\cdot 10^{-5}~{\rm pb}. 
\end{equation}
	The limitations of 
(\ref{DAMA-main-we}) should have some consequence for observables.
	Taking them into account we have obtained the 
	reduction of our scatter plots 
	for the total expected event rate of direct WIMP detection
	in a $^{73}$Ge detector
(Fig.~\ref{RateGe-damarg}) and
	the indirect detection rate for upgoing muons 
	from dark matter particles annihilation in the Earth and the Sun 
(Fig.~\ref{Upmuons-damarg}). 
	The calculations of 
	indirect detection rates follow the description given in 
\cite{Jungman:1996df,Bednyakov:2000uw}.
	There is also a reduction of allowed masses of some SUSY particles
(Fig.~\ref{Spectra-damarg}).
\begin{figure}[t!]  
\begin{picture}(100,167)
\put(-30,-33){\includegraphics{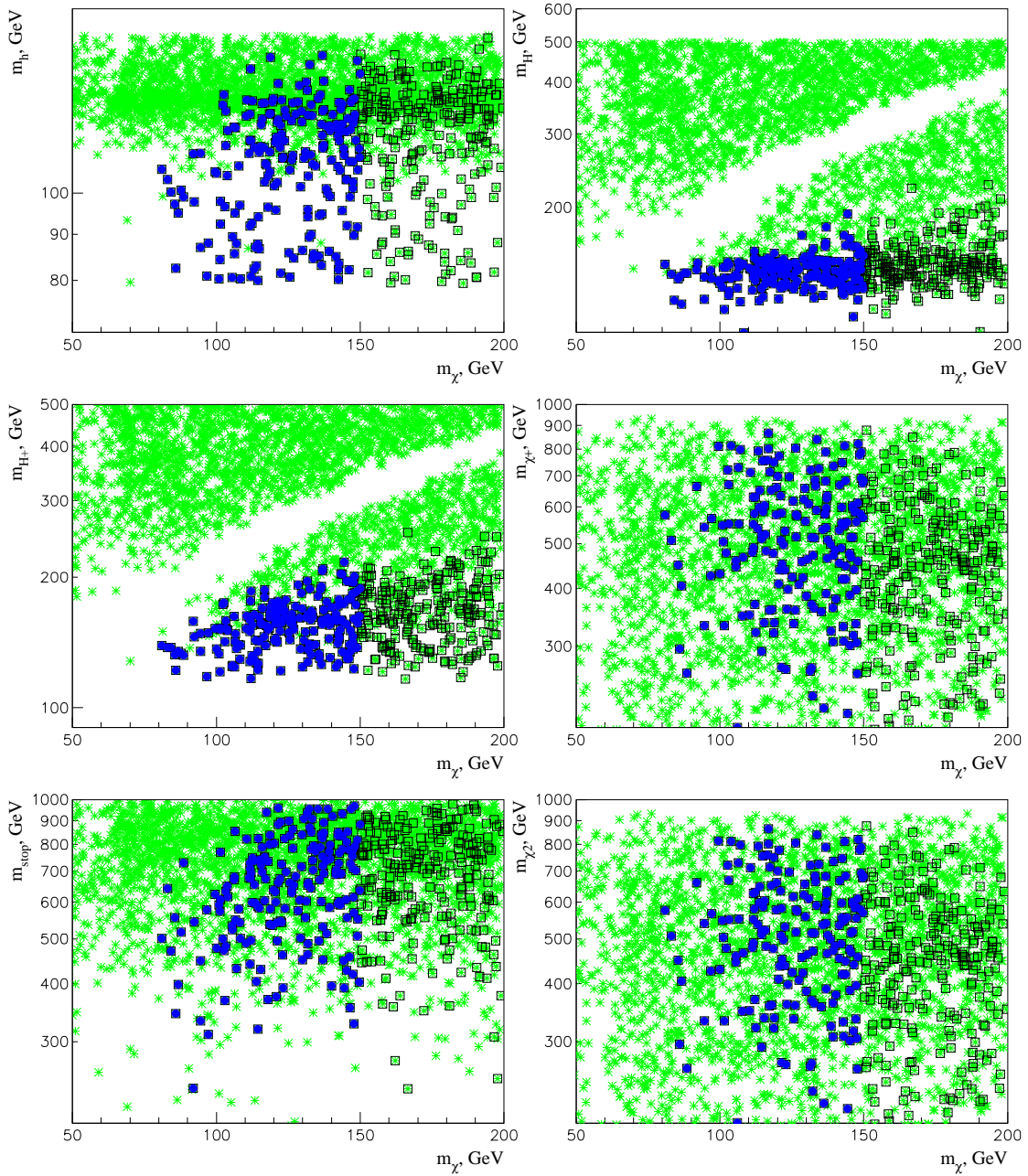}}
\end{picture}
\caption{Masses in GeV of light ($m^{}_{\rm h}$), 
        heavy ($m^{}_{\rm H}$), and charged Higgs bosons 
	($m^{}_{\rm H^+}$), 
	as well as masses of chargino ($m^{}_{\chi^+}$), 
	stop ($m^{}_{\rm stop}$), 
	and second neutralino ($m^{}_{\chi^2}$), 
	versus
	the mass ($m^{}_{\chi}$) of the LSP neutralino 
	under the same DAMA-inspired restrictions as in 
Figs.~\ref{RateGe-damarg},\ref{Upmuons-damarg}.
\label{Spectra-damarg}}
\end{figure} 
	In total from these figures one can see that
	the DAMA evidence favors the light Higgs sector of the MSSM, 
	relatively high event rate in Ge detectors, as well as 
	relatively high up-going muon fluxes from  the Earth and
	from the Sun for indirect detection of the relic neutralino. 
	It is also almost insensitive to the sfermion and 
	neutralino-chargino particle masses.
	As noted before in 
\cite{Bednyakov:2000uw,Bednyakov:1999vh}
	the relatively light Higgs masses (smaller than 200~GeV) 
	are very interesting from the point of accelerator
	SUSY searches.

\section{Conclusion} 
	In the effective low-energy MSSM (effMSSM) 
	for zero momentum transfer we calculated 
	the LSP-proton(neutron) 
	spin and scalar cross sections in the low LSP mass regime, 
	which follows from the DAMA dark matter evidence.
	We compared the calculated cross sections 
	with experimental exclusion
	curves and demonstrated that
        about a two-orders-of-magnitude improvement 
        of the current DM experiment sensitivities is needed 
        to reach the SUSY predictions for the $\sigma^{p,n}_{{\rm SD}}$. 

	We noted an in principle possible incorrectness 
	in the direct comparison of exclusion curves 
	for WIMP-proton(neutron) spin-dependent cross section 
	obtained with and without 
	non-zero WIMP-neutron(proton) spin-dependent contribution.
	On the other side, nuclear spin structure calculations
	show that usually one, WIMP-proton $\langle{\bf S}^A_{p}\rangle$, 
	or WIMP-neutron $\langle{\bf S}^A_{n}\rangle$, 
	nuclear spin dominates 
	and in the effMSSM
        we have the WIMP-proton and WIMP-neutron effective 
	couplings $a_{n}$ and $a_p$ of the same order of magnitude
(Fig.~\ref{an2ap-ratio}).
	Therefore at the current level
	of accuracy it looks reasonable to safely neglect 
	sub-dominant WIMP-nucleon contributions 
	analyzing the data from spin-non-zero targets.
	Furthermore the above-mentioned incorrectness concerns 
	also the direct comparison
	of spin-dependent exclusion curves obtained with and without non-zero
	spin-independent contributions
\cite{Bernabei:2003za,Bernabei:2003wy}.
	To be consistent,  for this comparison one has 
	to use a mixed spin-scalar coupling approach
(Figs.~\ref{Bernabei:2001ve:fig} and
       \ref{hdms-toy}), 
	as for the first time proposed by the DAMA collaboration
\cite{Bernabei:2000qi,Bernabei:2003za,Bernabei:2003wy}. 
	We applied such spin-scalar coupling approach to estimate 
	future prospects of the HDMS experiment with 
        the neutron-odd group high-spin isotope $^{73}$Ge. 
	Although the odd-neutron nuclei $^{73}$Ge, $^{129}${Xe} 
	already with the present accuracy lead to some sharper 
	restrictions for $\sigma^{n}_{\SD}$ then obtained by DAMA,  
	we found that the current accuracy of measurements with 
        $^{73}$Ge (as well as with $^{129}$Xe and NaI)
        did not yet reach a level which allows us to obtain 
        new decisive constraints on the SUSY parameters.
        Future about two-orders-of-magnitude 
	improvement of the background index 
        in the HDMS experiment
\cite{Klapdor-Kleingrothaus:2002pg} 
        can in principle 
        supply us with new constraints for the SUSY models.
	
	Finally we noticed that 
	the DAMA evidence favors the 
	light Higgs sector in the effMSSM (which could be reached at LHC), 
	a high event rate in a $^{73}${Ge} detector and 
	relatively high upgoing muon fluxes 
	from relic neutralino annihilations 
	in the Earth and the Sun. 

\enlargethispage{\baselineskip}

\smallskip 
        V.B. thanks the Max Planck Institut f\"ur Kernphysik 
        for the hospitality and RFBR (Grant 02--02--04009) for support.

\clearpage

{\small
\providecommand{\href}[2]{#2}\begingroup\raggedright\endgroup
 
}

\end{document}